\documentclass[a4paper,twoside,english,peerreview]{IEEEtran}
\usepackage[T1]{fontenc}
\usepackage[latin9]{inputenc}
\usepackage{fancyhdr}
\usepackage{graphicx}

\providecommand{\tabularnewline}{\\}

\usepackage{xcolor}
\usepackage{soul}

\usepackage{babel}

\begin{document}

\title{Stability Variances: A filter Approach. }

\author{Alaa~Makdissi, François~Vernotte and~Emeric~De Clercq%
\thanks{A. Makdissi is with ELA MEDICAL (SORIN Group), C.A. La Boursidière,
F-92357 Le Plesis Robinson, \texttt{http:$//$www.alamath.com}, %
}%
\thanks{F. Vernotte is with Institut Utinam, UMR CNRS 6213, Observatoire de
Besançon, Université de Franche-Comté, 41~bis~avenue de l'observatoire,
BP~1615, F-25010~Besançon~cedex.%
}%
\thanks{E. De Clercq is with LNE-SYRTE, UMR CNRS 8630, Observatoire de Paris,
61~avenue de l'observatoire, F-75014~Paris%
}}

\maketitle
\fancyhead{} \fancyhead[LE]{IEEE Transactions on Ultrasonics, Ferroelectrics, and Frequency Control, 2009}
\fancyhead[LO]{Makdissi \MakeLowercase{\textit{et al.}}: Stability Variances: A filter Approach}

\bibliographystyle{IEEEtran}
\begin{abstract}
We analyze the Allan Variance estimator as the combination of Discrete-Time
linear filters. We apply this analysis to the different variants of
the Allan Variance: the Overlapping Allan Variance, the Modified Allan
variance, the Hadamard Variance and the Overlapping Hadamard variance.
Based on this analysis we present a new method to compute a new estimator
of the Allan Variance and its variants in the frequency domain. We
show that the proposed frequency domain equations are equivalent to
extending the data by periodization in the time domain. Like the Total
Variance \cite{totvar}, which is based on extending the data manually
in the time domain, our frequency domain variances estimators have
better statistics than the estimators of the classical variances in
the time domain. We demonstrate that the previous well-know equation
that relates the Allan Variance to the Power Spectrum Density (PSD)
of continuous-time signals is not valid for real world discrete-time
measurements and we propose a new equation that relates the Allan
Variance to the PSD of the discrete-time signals and that allows to
compute the Allan variance and its different variants in the frequency
domain . 
\end{abstract}
\IEEEpeerreviewmaketitle

\section{Introduction}

The Allan Variance \cite{allan} and other frequency stability variances
\cite{barnes,linchi,modallan,totvar} were introduced in order to
allow characterization and classification of frequency fluctuations
\cite{rutman}. One of the goals of these frequency stability variances
was to overcome the fact that the true variance is mathematically
undefined in the case of some power law spectrum \cite{rutman}.

The stability properties of oscillators and frequency standards can
be characterized by two ways: the power spectral density (PSD) of
the phase (or frequency) fluctuations, i. e. the energy distribution
in the Fourier frequency spectrum; or various variances of the frequency
fluctuations averaged during a given time interval, it is said in
the time domain. The power spectral density of frequency fluctuations
is of great importance because it carries more information than the
time domain frequency stability variances and provides an unambiguous
identification of the noise process encountered in real oscillators.
PSD are the preferred tool in several applications such as telecommunications
or frequency synthesis. Stability variances are most used in systems
in which time measurements are involved, or for very low Fourier frequencies.
Each one of these tools corresponds to a specific instrumentation,
spectrum analyzers for frequency-domain measurements, and digital
counters for time domain measurements. Although there is a separation
between measurements methods, use and sometimes user's community of
these two parameters, time-domain and frequency-domain parameters
naturally are not independent. The true variance for example can be
theoretically deduced from the PSD by an integral relationship. The
true variance $\sigma_{Y}^{2}$ of a zero-mean continuous-time signal
$Y(t)$ is defined for stationary signals as the value of the autocorrelation
function $R_{Y}\left(\tau\right)=E\left[Y\left(t\right)Y\left(t+\tau\right)\right]$
for $\tau=0$ (where $E$ is the mathematical expectation operator)
\cite{roddier}. This statistical definition of the autocorrelation
is related to the time-averge of the product $Y\left(t\right)Y\left(t+\tau\right)$
if the signal is correlation-ergodic \cite{papoulis} by:

\begin{equation}
R_{Y}(\tau)=\lim_{T\to\infty}\frac{1}{2T}\int\limits _{-T}^{T}Y\left(t+\tau\right)Y(t)dt\label{eq:0}\end{equation}

The definition of the two-sided Power Spectral Density (PSD) $S_{Y}^{TS}\left(f\right)$
of the signal \emph{Y} is related to Autocorrelation function by the
Fourier Transform and its inverse by \cite{roddier}: \begin{equation}
S_{Y}^{TS}\left(f\right)=\int\limits _{-\infty}^{\infty}R_{Y}\left(\tau\right)e^{-i2\pi f\tau}d\tau\label{eq:1}\end{equation}
 and \begin{equation}
R_{Y}\left(\tau\right)=\int\limits _{-\infty}^{\infty}S_{Y}^{TS}\left(f\right)e^{i2\pi f\tau}df.\label{eq:2}\end{equation}

The two-sided PSD is a positive ($S^{TS}(f)>0$) and a symetric function
in $f$ ($S^{TS}(f)=S^{TS}(-f)$ ). In frequency metrology, the single-sided
Power Spectral Density $S_{Y}(f)$ has been historically utilized.
It is related to the two-sided PSD by :

\begin{equation}
S_{Y}(f)=\left\{ \begin{array}{lcc}
2S_{Y}^{TS}(f) & \textrm{if} & f\geq0\\
{\displaystyle 0} & \textrm{if} & f<0.\end{array}\right.\label{eq:3}\end{equation}

For power-law spectrum signals, the PSD is expressed as $S_{Y}\left(f\right)=h_{\alpha}f^{\alpha}$
\cite{rutman}. The $\alpha$ integer value may vary from -4 to +2
in common clocks frequency fluctuation signals \cite{howe}. The true
variance is defined then as \cite{rutman}: \begin{equation}
\sigma_{Y}^{2}=R_{Y}\left(0\right)=\int\limits _{0}^{\infty}S_{Y}\left(f\right)df=\int\limits _{0}^{\infty}h_{\alpha}f^{\alpha}df.\label{eq:varvrai}\end{equation}

We can notice easily that for integer $\alpha<0$, $\lim_{f\to0}f^{\alpha}$
diverges and then the integral in (\ref{eq:varvrai}) is infinite.

The intent of this paper is to explore the relationship between stability
variances and the PSD using a filter approach. This approach allows
us to establish new estimators of the classical known variances (Allan,
Hadamard) in the frequency domain instead of the time domain, especially
in the case of discrete signals, which are the most current in practice.
The filter approach analysis is developed in Section II in the general
case of a difference filter of order \emph{n}. This approach allows
us to propose general formulae for the stability variance of continuous-time
signals. The well known frequency stability variances like (AVAR,
MODAVAR, HADAMARD) are special cases of the proposed formula for \emph{n}=1
and \emph{n}=2. As in practical application the signals are not continuous
because the measurement instruments are read at discrete periodic
instants, the filter approach is then extended in Section III to discrete-time
signals. New estimators of the classical variances in the frequency
domain are proposed which are different from a simple discretization
of the integral of the continuous-time equations. The proposed discrete-time
variances are based on the fact that filtering in the discrete frequency
domain is equivalent to a periodization in the time domain. This periodization
makes our proposed variances estimator have a better statistics than
the classical estimators. In Section IV we present the theoretical
calculation of the equivalent degree of freedom of the new proposed
frequency domain variances estimators. Finally, these estimators,
the overlapping Allan variance (OAVAR), the Hadamard variance (HVAR),
and the modified Allan variance (MAVAR), are compared in Section V
to the same estimators in the time-domain using a numerical simulation.

\section{Continous-time signals}

\subsection{Characterization of long term stability by filtering}

\label{sect:ContiVAR}Often, it's desirable to characterize the long
term stability of clocks. Long term behaviour is determined by the
components of the PSD at low frequencies ($f$ tends to zero). In
order to obtain the long term behaviour we average the signal $Y(t)$
and we study the variance of the averaged signal. Let $Z(t)$ the
signal obtained by averaging the signal $Y(t)$ during a time $\tau$.
We can write then: \begin{equation}
Z\left(t,\tau\right)=\frac{1}{\tau}\int_{t-\tau}^{t}Y\left(t\right)dt.\label{eq:4}\end{equation}

The signal $Z\left(t,\tau\right)$ could be seen as the output of
a moving average filter $M$ of length $\tau$. The moving average
filter impulse response ${m\left(t,\tau\right)}$ is defined by: \begin{equation}
m\left(t,\tau\right)=\frac{1}{\tau}\textrm{Rect}_{\tau}\left(t-\frac{\tau}{2}\right),\label{eq:5}\end{equation}
 where $\textrm{Rect}_{B}\left(t\right)$ is a centered rectangular
windows of width $B$: \begin{equation}
\textrm{Rect}_{B}\left(t\right)=\left\lbrace \begin{array}{ll}
1 & \textrm{for}\qquad-\frac{B}{2}\le t\le\frac{B}{2}\\
0 & \textrm{otherwise.}\end{array}\right.\label{eq:6}\end{equation}

Thus, in the time domain, $Z(t,\tau)$ may be defined as: \begin{equation}
Z(t,\tau)=m(t,\tau)\ast Y(t),\label{eq:rajout}\end{equation}
 where `$\ast$' denotes the convolution product operator.

The frequency Response $M(f)$ of this moving average filter is given
by: \begin{equation}
M\left(f\right)=\frac{\sin\left(\pi\tau f\right)}{\pi\tau f}e^{i\pi\tau f}.\label{eq:mov-ave}\end{equation}

According to linear filter properties, the PSD $S_{Z}\left(f\right)$
of the continuous time signal $Z\left(t,\tau\right)$ is: \begin{equation}
S_{Z}\left(f\right)=\left|M\left(f\right)\right|^{2}S_{Y}\left(f\right).\label{eq:8}\end{equation}

From (\ref{eq:varvrai}) (\ref{eq:mov-ave}) and (\ref{eq:8}), the
variance of the $Z\left(t,\tau\right)$ signal is expressed by: \begin{equation}
\sigma_{Z}^{2}\left(\tau\right)=\int_{0}^{\infty}\frac{\sin^{2}\left(\pi\tau f\right)}{\left(\pi\tau f\right)^{2}}S_{Y}\left(f\right)df.\label{eq:defsigz}\end{equation}

It's clear that when $S_{Y}\left(f\right)=h_{\alpha}f^{\alpha}$ the
variance $\sigma_{Z}^{2}\left(\tau\right)$ is not defined for power
law with $\alpha<0$ because the $M(f)$ filter tends to 1 when $f$
tends to zero. In order to make the variance $\sigma_{Z}^{2}\left(\tau\right)$
defined when $\alpha<0$ we need to introduce an additional filter
$D(f)$ in series with $M(f)$. The input of the new $D(f)$ filter
is $Z\left(t,\tau\right)$ and let us call its output $U\left(t,\tau\right)$.
The variance of the $U\left(t,\tau\right)$ signal is expressed when
$Y(t)$ has power law spectrum by: \begin{equation}
\sigma_{U}^{2}\left(\tau\right)=h_{\alpha}\int_{0}^{\infty}\frac{\sin^{2}\left(\pi\tau f\right)}{\left(\pi\tau f\right)^{2}}\left|D\left(f\right)\right|^{2}f^{\alpha}df.\label{eq:10}\end{equation}

Obviously, the variance $\sigma_{U}^{2}\left(\tau\right)$ becomes
defined if $\lim_{f\to0}\left|D\left(f\right)\right|^{2}f^{\alpha}$
is defined. This means that $D(f)$ must be of the form $f^{\beta}$
when $f\rightarrow0$ with $\beta>-\alpha/2$. In common clock noise
with $-4\le\alpha\le+2$ the filter $D(f)$ must verifies $D\left(f\right)\propto f^{2}$
approximately for sufficiently small $f$ in order to make $\sigma_{U}^{2}\left(\tau\right)$
defined for the seven common clock noises.

This processing may seem contradictory in the sense that we are looking
for the long term behaviour (i.e when $f\rightarrow0$) of the signal
$Y(t)$ and the proposed processing introduces in the same time a
filter $D(f)$ that eliminates to a certain extent the components
of the PSD of $Y(t)$ at $f=0$. In fact, even if the introduced processing
may cancel the component of $S_{Y}\left(f\right)$ at $f=0$ and hence
makes tend $\sigma_{U}^{2}\left(\tau\right)$\
to 0 when $\tau\rightarrow\infty$, such a processing allows to study
the asymptotic behaviour of $S_{Y}\left(f\right)$ when $f$ approaches
zero. We will see in the following of this paper that this asymptotic
behaviour allows to characterize and classify the noise signals.

In order to realize a filter $D(f)$ with a frequency response $D\left(f\right)=f^{\beta}$,
the first idea that comes to mind is to use multiple continuous time
derivations of the signal $Z\left(t,\tau\right)$. Each derivation
in the time domain is equivalent to a multiplication by $f^{2}$ in
the PSD domain.

Let $U^{(n)}\left(t,\tau\right)$ the $n^{\textrm{\footnotesize th}}$
derivative of $Z\left(t,\tau\right)$ defined by: \begin{equation}
U^{(n)}\left(t,\tau\right)=\frac{d^{n}Z\left(t,\tau\right)}{dt^{n}}.\label{eq:11}\end{equation}

The PSD $S_{U}^{(n)}\left(f\right)$ of $U^{(n)}\left(t,\tau\right)$
is given by: \begin{equation}
S_{U}^{(n)}\left(f\right)=\left(2\pi f\right)^{2n}S_{Z}\left(f\right)=\left(2\pi f\right)^{2n}\frac{\sin^{2}\left(\pi\tau f\right)}{\left(\pi\tau f\right)^{2}}S_{Y}\left(f\right),\label{eq:12}\end{equation}
 and its variance $\sigma_{U}^{2}\left(\tau,n\right)$ is equal to:
\begin{equation}
\sigma_{U}^{2}\left(\tau,n\right)=\int_{0}^{\infty}\left(2\pi f\right)^{2n}\frac{\sin^{2}\left(\pi\tau f\right)}{\left(\pi\tau f\right)^{2}}S_{Y}\left(f\right)df.\label{eq:defsigun}\end{equation}

A perfect continuous derivation in the time domain has a linear frequency
response for all the frequencies. Such a derivation is impossible
to realize and is often approximated by a filter $D$ that have the
same frequency response in the vicinity of $f=0$. The simplest filter
that approximates a derivation is the simple time difference filter
defined by its impulse response: \begin{equation}
d\left(t,\tau\right)=\delta\left(t\right)-\delta\left(t-\tau\right).\label{eq:defdttau}\end{equation}

Its Fourier Transform $D(f)$ is given by: \begin{equation}
D\left(f\right)=1-e^{-2i\pi\tau f}=2i\sin\left(\pi\tau f\right)e^{-i\pi\tau f}.\label{eq:derfilt}\end{equation}

When cascading $n$ simple difference filters $d\left(t,\tau\right)$,
we obtain a $n$-order difference filter. Its impulse response $d^{(n)}\left(t,\tau\right)$
is given by: \begin{equation}
d^{(n)}\left(t,\tau\right)=\sum_{k=0}^{n}\left(-1\right)^{k}C_{n}^{k}\delta\left(t-k\tau\right)\label{eq:defdnttau}\end{equation}
 where $C_{n}^{k}$ in the above equation is the binomial coefficient
defined by $C_{n}^{k}=\frac{n!}{k!\left(n-k\right)!}$ and $n!$ denotes
the factorial of $n$.

According to equation (\ref{eq:derfilt}), the frequency response
$D^{(n)}\left(f\right)$ of the $d^{(n)}\left(t,\tau\right)$ filter
is given by: \begin{equation}
D^{(n)}\left(f\right)=\left(D\left(f\right)\right)^{n}=2^{n}i^{n}\:\sin^{n}\left(\pi\tau f\right)e^{-i\pi n\tau f}.\label{eq:nth-der}\end{equation}

We choose to normalize this filter in such a way that it does not
modify the variance of a white noise processed by it. The normalization
factor $c_{n}$ is given by the square root of the sum of the squares
of the coefficients $\left(-1\right)^{k}C_{n}^{k}$: \begin{equation}
c_{n}^{2}=\sum_{k=0}^{n}\left[\left(-1\right)^{k}C_{n}^{k}\right]^{2}=\frac{4^{n}\Gamma\left(n+1/2\right)}{\sqrt{\pi}\Gamma\left(n+1\right)}=C_{2n}^{n}.\label{eq:coefcn}\end{equation}

The output $U^{(n)}\left(t,\tau\right)$ of the normalized filter
$d^{(n)}\left(t,\tau\right)/c_{n}$ is given by: \begin{equation}
U^{(n)}\left(t,\tau\right)=\frac{1}{c_{n}}\left[d^{(n)}\left(t,\tau\right)\ast m\left(t,\tau\right)\right]\ast Y\left(t\right).\label{eq:defUttau}\end{equation}

The variance of $U^{(n)}\left(t,\tau\right)$ is expressed by: \begin{equation}
\sigma_{U}^{2}\left(\tau\right)_{(n)}=\frac{1}{c_{n}^{2}}\int_{0}^{\infty}\left|D^{(n)}\left(f\right)M\left(f\right)\right|^{2}S_{Y}(f)df\label{eq:20}\end{equation}
 or, equivalently, using equations (\ref{eq:mov-ave}) and (\ref{eq:nth-der}):
\begin{equation}
\sigma_{U}^{2}\left(\tau\right)_{(n)}=\frac{2^{2n}}{c_{n}^{2}}\int_{0}^{\infty}\frac{\sin^{{2n+2}}\left(\pi\tau f\right)}{\left(\pi\tau f\right)^{2}}S_{Y}(f)df.\label{eq:varUspec}\end{equation}

The convergence domain of this variance is given by $\alpha>-\left(n+2\right)$.
For positive $\alpha$ values we must introduce a high cut-off frequency
as the upper limit of the integration in order to insure the convergence
of $\sigma_{U}^{2}\left(\tau\right)_{(n)}$.

Sometimes it's useful to express the variance $\sigma_{U}^{2}\left(\tau\right)_{(n)}$
versus the PSD ${S_{X}\left(f\right)}$ of the phase signal $X(t)$
related to the frequency fluctuation $Y(t)$ by $Y(t)=\frac{dX(t)}{dt}$.
Replacing $S_{Y}\left(f\right)=\left(2\pi f\right)^{2}S_{X}(f)$ in
(\ref{eq:varUspec}) we get: \begin{equation}
\sigma_{U}^{2}\left(\tau\right)_{(n)}=\frac{2^{2n+2}}{c_{n}^{2}\tau^{2}}\int_{0}^{\infty}\sin^{2n+2}\left(\pi\tau f\right)S_{{X}}\left(f\right)df.\label{eq:22}\end{equation}

Thus, according to the order $n$ of the used difference filter $d^{(n)}\left(t,\tau\right)$
we obtain different variances with different convergence domains (see
\cite{linchi} and \cite{vernotte02c} for the explicit link between
$n$ and the convergence). We will see in the next of this paper that
most of the well-known stability variances are special cases of equation
(\ref{eq:varUspec}) or (\ref{eq:22}).

\subsection{The Allan Variance and the Hadamard Variance as filters}

When the order $n$ of the filter $d^{(n)}\left(t,\tau\right)$ is
equal to one, $c_{1}^{2}=2$ and, from (\ref{eq:varUspec}), we obtain
the Allan Variance defined by: \begin{equation}
\sigma_{y}^{2}\left(\tau\right)=\sigma_{U}^{2}\left(\tau\right)_{(1)}=2\int_{0}^{\infty}\frac{\sin^{4}\left(\pi\tau f\right)}{\left(\pi\tau f\right)^{2}}S_{Y}\left(f\right)df.\label{eq:filavar}\end{equation}

The Allan Variance is noted $\sigma_{y}^{2}\left(\tau\right)$ in
the literature but it's the true variance of $U^{(1)}\left(t,\tau\right)$,
a version of \emph{Y(t)} processed by filters $M$ and $D$.

Equation (\ref{eq:filavar}) shows that the Allan variance is defined
for power law spectrum with $\alpha$ values from -2 to 0. For $\alpha>0$,
the Allan variance does not converge unless a high cut-off frequency
$f_{h}$ is taken into account. Moreover the asymptotic behaviour
of $\sigma_{U}^{2}\left(\tau\right)$ is similar for the White Phase
noise ($\alpha=2$) and Flicker Phase Noise ($\alpha=1$) (see table
\ref{tab:ahvar}). For power law with $\alpha=-3$ and $\alpha=-4$
the Allan variance is undefined (unless a low cut-off frequency is
taken into account).

When the order $n$ of the filter $d^{(n)}\left(t,\tau\right)$ is
equal to 2, $c_{2}^{2}=6$ and we obtain the three sample Hadamard
variance \cite{hadamard} also called the Picinbono variance \cite{picinbono}.
From (\ref{eq:varUspec}), this variance is defined by: \begin{equation}
\sigma_{H}^{2}\left(\tau\right)=\sigma_{U}^{2}\left(\tau\right)_{(2)}=\frac{8}{3}\int_{0}^{\infty}\frac{\sin^{6}\left(\pi\tau f\right)}{\left(\pi\tau f\right)^{2}}S_{Y}\left(f\right)df.\label{eq:filhadam}\end{equation}

This equation shows that the Hadamard variance is defined for law
power spectrum with integer $\alpha$ values between -4 and 0. As
previously explained, a high cut-off frequency $f_{h}$ is necessary
for $\alpha>0$ in order to ensure convergence of the integral (\ref{eq:filhadam})
when $f\rightarrow\infty$.

Table \ref{tab:ahvar} shows the values of Allan variance \cite{rutman}
and Hadamard variance \cite{hadamard,picinbono,Walter} for power
law spectra. The results reported in this table if $\alpha>0$ are
only valid for $\tau\gg1/\left(2\pi f_{h}\right)$.

\begin{table*}
\centering 

\caption{Allan and Hadamard variances for power law spectra. $\gamma\approx0.577216$
is the Euler's constant and $f_{h}$ is the high cut-off frequency
for noise with $\alpha>0$.}

\begin{tabular}{|c|c|c|}
\hline 
\boldmath{$\alpha$}  & \textbf{Allan Variance} \boldmath{$\sigma_{y}^{2}\left(\tau\right)$}  & \textbf{Hadamard Variance} \boldmath{$\sigma_{H}^{2}\left(\tau\right)$}\tabularnewline
\hline 
 &  & \tabularnewline
+2  & ${\displaystyle \frac{3f_{h}}{4\pi^{2}\tau^{2}}h_{+2}}$  & ${\displaystyle \frac{5f_{h}}{6\pi^{2}\tau^{2}}h_{+2}}$ \tabularnewline
 &  & \tabularnewline
\hline 
 &  & \tabularnewline
+1  & ${\displaystyle \frac{3\left[\gamma+\ln\left(2\pi f_{h}\tau\right)\right]-\ln\left(2\right)}{4\pi^{2}\tau^{2}}h_{+1}}$  & ${\displaystyle \frac{10\left[\gamma+\ln\left(2\pi f_{h}\tau\right)\right]+\ln\left(3\right)-\ln\left(64\right)}{12\pi^{2}\tau^{2}}h_{+1}}$ \tabularnewline
 &  & \tabularnewline
\hline 
 &  & \tabularnewline
0  & ${\displaystyle \frac{1}{2\tau}h_{0}}$  & ${\displaystyle \frac{1}{2\tau}h_{0}}$ \tabularnewline
 &  & \tabularnewline
\hline 
 &  & \tabularnewline
-1  & ${\displaystyle 2\ln\left(2\right)h_{-1}}$  & ${\displaystyle \frac{1}{2}\ln\left(\frac{256}{27}\right)h_{-1}}$ \tabularnewline
 &  & \tabularnewline
\hline 
 &  & \tabularnewline
-2  & ${\displaystyle \frac{2\pi^{2}\tau}{3}h_{-2}}$  & ${\displaystyle \frac{\pi^{2}\tau}{3}h_{-2}}$ \tabularnewline
 &  & \tabularnewline
\hline 
 &  & \tabularnewline
-3  & --  & ${\displaystyle \frac{8\pi^{2}\tau^{2}}{3}\left[\frac{27}{16}\ln\left(3\right)-\ln\left(4\right)\right]h_{-3}}$ \tabularnewline
 &  & \tabularnewline
\hline 
 &  & \tabularnewline
-4  & --  & ${\displaystyle \frac{11\pi^{4}\tau^{3}}{15}h_{-4}}$ \tabularnewline
 &  & \tabularnewline
\hline
\end{tabular}\label{tab:ahvar} 
\end{table*}

Because $D\left(f\right)\propto f^{n}$ in the vicinity of zero we
can say that $D$-filtering is equivalent to high-pass filtering.
The combination of the low pass filter $M(f)$ with the high-pass
filter $D(f)$ forms a band-pass filter $G(f)$. We will see in the
following of this paper that all the stability variances could be
expressed as the variance of output of band-pass filters applied to
the signal under study $Y(t)$. When varying $\tau$ we obtain different
band-pass filters (a filter bank) with different bandwidths. This
analysis is similar to the multi-resolution wavelet analysis \cite{vernotte02c}
and the special case of the Allan variance filter is nothing else
but the Haar wavelet basis function \cite{percival}.

It's worth recalling that equation (\ref{eq:filavar}) is valid only
for continuous time signal and filters. This equation gives a theoretical
definition of the Allan Variance of the continuous signal $Y(t)$
and cann't be used to compute the Allan variance unless the formal
expression of the PSD $S_{Y}(f)$ is a known function. In real world
application signals are collected at discrete instants and the above
$M$ and $D$ filters are unrealizable for big values of $\tau$ especially
when $\tau$ duration may last for months and years. In the next section
we analyse the stability variances in the case of discrete-time signals.

\section{Discrete-time variances}

In real world applications, measurement instruments are read at discrete
periodic instants. Let $T$ be the period of the reading cycle. We
suppose that the instrument measures the mean value during this cycle
without dead time. We have then a discrete time series or signal given
by: \begin{equation}
y_{k}=\frac{1}{T}\int_{(k-1)T}^{kT}Y\left(t\right)dt.\label{eq:def_yk}\end{equation}

The time-series $y_{k}$ of a finite length is converted to digital
numbers and is studied in order to characterize and classify the continuous
time signal $Y(t)$. The PSD $S_{y}\left(f\right)$ of the discrete-time
signal is periodic with a period $f_{s}=1/T$ and is related to the
PSD $S_{Y}\left(f\right)$ of the continuous signal $Y(t)$ by: \begin{equation}
S_{y}\left(f\right)=\frac{1}{T}\sum_{n}S_{Y}\left(f-nf_{s}\right)\frac{\sin^{2}\left[\pi T\left(f-nf_{s}\right)\right]}{\left[\pi T\left(f-nf_{s}\right)\right]^{2}}.\label{eq:PSDper}\end{equation}

We notice from equation (\ref{eq:PSDper}) that the PSD $S_{y}(f)$
is equal to $S_{Y}(f)/T$ when $f\rightarrow0$ because all the terms
in the sum are null ($\sin\left(n\pi\right)=0$) except the term for
$n=0$. We conclude that we can study the long term behaviour of the
continuous signal $Y(t)$ by using the discrete time series $y_{k}$.
We can show without difficulty that in the presence of a dead-time
(sampling period larger than the averaging period) we have an aliasing
phenomenon even for $f\rightarrow0$.

In some applications it's possible to eliminate or reduce the aliasing
phenomenon by using a low pass filter inside the measurement instrument
in front of the moving average operation.

For frequencies varying between 0 and $f_{s}/2$, we can expect that
the PSD $S_{y}(f)$ of the discrete sequence $y_{k}$ is nearly equal
to $S_{Y}(f)/T$, at least in the case of a white noise, because averaging
during a time $T$ and then sampling with a period $T$ preserve most
of the information contained in the signal $Y(t)$, since the averaging
can be considered as a non perfect anti-aliasing low pass filter.

For power-law spectrum the sum in equation (\ref{eq:PSDper}) can
be expressed formally for $\alpha<0$. For $\alpha>0$, we must introduce
a high cut-off frequency $f_{h}$. Table \ref{tab:PSDdisc} shows
the expression of $S_{y}(f)$ for some negative $\alpha$ values when
$f$ varies between 0 and $f_{s}/2$. The formulae in Table \ref{tab:PSDdisc}
relating the PSD of the sampled signal to the PSD of the continuous
signal were never published before to our best knowledge.

\begin{table}
\centering 

\caption{The PSD $S_{y}(f)$ of the sampled time series $y_{k}$ when $Y(t)$
has a power law spectrum $S_{Y}(f)=h_{\alpha}f^{\alpha}$. \quad{}$\psi(n,x)$
is the Polygamma function defined by $\psi(n,x)=(-1)^{n+1}n!\sum_{k=0}^{\infty}1/(x+k)^{n+1}$.}

\begin{tabular}{|c|c|}
\hline 
\boldmath{$\alpha$}  & \boldmath{$TS_{y}(f)$} \tabularnewline
\hline 
 & \tabularnewline
0  & $h_{0}$ \tabularnewline
 & \tabularnewline
\hline 
 & \tabularnewline
-1  & ${\displaystyle h_{-1}\frac{\left[1-T^{3}f^{3}\psi\left(2,1+fT\right)\right]\sin^{2}\left(\pi fT\right)}{\pi^{2}T^{2}f^{3}}}$ \tabularnewline
 & \tabularnewline
\hline 
 & \tabularnewline
-2  & ${\displaystyle h_{-2}\frac{\pi^{2}T^{2}}{3}\frac{\left[2+\cos\left(2\pi fT\right)\right]}{\sin^{2}\left(\pi fT\right)}}$ \tabularnewline
 & \tabularnewline
\hline 
 & \tabularnewline
-3  & ${\displaystyle h_{-3}\frac{\left[12-T^{5}f^{5}\psi\left(4,1+fT\right)\right]\sin^{2}\left(\pi fT\right)}{12\pi^{2}T^{2}f^{5}}}$ \tabularnewline
 & \tabularnewline
\hline 
 & \tabularnewline
-4  & ${\displaystyle h_{-4}\frac{\pi^{4}T^{4}}{60}\frac{\left[33+26\cos\left(2\pi fT\right)+\cos\left(4\pi fT\right)\right]}{\sin^{4}\left(\pi fT\right)}}$ \tabularnewline
 & \tabularnewline
\hline
\end{tabular}\label{tab:PSDdisc} 
\end{table}

A Taylor expansion of $S_{y}(f)$ when $f$ tends to zero ($\alpha<0$)
gives (see table \ref{tab:PSDdisc}): \begin{equation}
S_{y}\left(f\right)=\frac{1}{T}\left[h_{\alpha}f^{\alpha}-h_{\alpha}\frac{\pi^{2}T^{2}}{3}f^{\alpha+2}\right]=\frac{S_{Y}(f)+A(f)}{T}.\label{eq:PSDneg}\end{equation}

We call $A(f)=-h_{\alpha}\frac{\pi^{2}T^{2}}{3}f^{\alpha+2}$ the
aliasing term for integer $\alpha<0$. It depends on the sampling
period $T$ and is null for white noise ($\alpha=0$). At long term,
the dominant component in (\ref{eq:PSDneg}) is $S_{Y}(f)/T$ and
the aliasing is negligible. For short term ($f\rightarrow\frac{1}{2T}$)
the aliasing term varies as $T^{-\alpha}$ and increases when the
sampling period grows.

For $\alpha>0$, the aliasing term depends also on the high cut-off
frequency and varies as $f^{2}$ whatever the value of $\alpha$.
This means that the study of the stability variance of $y_{k}$ for
power-law spectra with $\alpha>2$ does not allow to study the behaviour
of $Y(t)$ because the aliasing term is dominant when $f$ tends to
zero \cite{metro98,vernotte98b}.

The variance $\sigma_{y}^{2}$ of the discrete time series $y_{k}$
is related to its periodic PSD $S_{y}(f)$ by \cite{papoulis}: \begin{equation}
\sigma_{y}^{2}=T\int_{-\frac{1}{2T}}^{+\frac{1}{2T}}S_{y}^{TS}(f)df.\label{eq:sigy_sy}\end{equation}

Equation (\ref{eq:PSDneg}) relates the PSD of the measured discrete
time signal $y_{k}$ (after averaging without dead-time) to the PSD
of the continuous-time signal $Y(t)$. Equation ( \ref{eq:sigy_sy}
) relates the variance to the PSD of the discrete-time signal. Combining
theses two equations and using an approach similar to that presented
in paragraph ( \ref{sect:ContiVAR} ) in the case of general difference
filter of order $n$ for the continuous-time signals, allows us to
define a general stability variance for discrete-time signal similar
to that of equation ( \ref{eq:varUspec} ) for continuous-time signals. 

In the case of a frequency fluctuation sequence, the time series $y_{k}$
could be related to the time error samples $X(t)$ by: \begin{equation}
y_{k}=\frac{1}{T}\int_{(k-1)T}^{kT}\frac{dX(t)}{dt}dt=\frac{X\left[kT\right]-X\left[\left(k-1\right)T\right]}{T}.\label{eq:yk_xt}\end{equation}

Sometimes, it's difficult to realize experimentally the measurement
of $y_{k}$ according to equation (\ref{eq:def_yk}) by averaging
and recording $y_{k}$ without dead-time. If the time error data $X(t)$
are measurable it is always possible to sample them and compute $y_{k}$
according to equation (\ref{eq:yk_xt}) without dead-time.

In order to simplify notations, we suppose, without loss in generality,
that $T$ is equal to 1 in the following of the paper. Then, integration
in equation (\ref{eq:sigy_sy}) is done over the interval $\left[-1/2,1/2\right]$
and equation (\ref{eq:yk_xt}) could be written, by denoting $x_{k}=X\left(kT\right)$,
as: \begin{equation}
y_{k}=x_{k}-x_{k-1}.\label{eq:30}\end{equation}

In other terms, the time error sequence $x_{k}$ could be obtained
from the averaged frequency signal $y_{k}$ by numerical integration
with a starting point $x_{0}=0$: \begin{equation}
x_{k+1}=x_{k}+y_{k}.\label{eq:31}\end{equation}

In order to estimate the $\sigma_{U}^{2}(\tau)_{(n)}$ variance from
the observed discrete-time series ${y_{k}}$ we try to realize a discrete
version ${u_{k}}$ of the continuous signal $U^{(n)}(t,\tau)$ defined
by equation (\ref{eq:defUttau}) by using digital filters similar
to the analog filters $m(t,\tau)$ and $d^{(n)}(t,\tau)$. Once we
have a discrete version of $U^{(n)}(t,\tau)$, we can estimate its
variance by computing the sample variance of the discrete-time series
$u_{k}$.

Following the filter approach used for continuous time signals we
introduce digital filters in such a way that their discrete-time outputs
are similar, as much as possible, to analog signals in the previous
section.

The moving average filter $m(t,\tau)$ of length $\tau$ becomes in
the discrete domain a rectangular windows of length $m=\tau/T$. The
output $z_{k}$ of this filter is given by: \begin{equation}
z_{k}=\frac{1}{m}\sum_{n=0}^{m-1}y_{k-n}.\label{eq:def_zk}\end{equation}

Obviously, by using (\ref{eq:def_zk}) and (\ref{eq:def_yk}), we
can write: \begin{equation}
z_{k}=\frac{1}{mT}\int_{(k-m)T}^{kT}Y\left(t\right)dt.\label{eq:defzcont}\end{equation}

{Equation (\ref{eq:defzcont}) shows that averaging $m$ values of
the signal } ${y_{k}}${ is equivalent to using an instrument with
an averaging time } ${\tau={mT}}${. This may let us think wrongly
that the PSD } ${S_{{z}}\left(f\right)}${, of the discrete time
series } ${z_{k}}${ could be obtained directly from equation (\ref{eq:8})
by replacing } ${\tau={mT}}$.

In fact, ${z_{k}}${ being discrete, its PSD is periodic and contains
aliasing terms. The PSD } ${S_{{z}}\left(f\right)}${ of the discrete
time series } ${z_{k}}${ is related to the PSD } ${S_{Y}\left(f\right)}${
of the continuous signal \emph{Y(t) }by:} \begin{equation}
S_{{z}}\left(f\right)=\frac{1}{T}\sum_{n}S_{Y}\left(f-nf_{s}\right)\frac{\sin^{2}\left[\pi mT\left(f-nf_{s}\right)\right]}{\left[\pi mT\left(f-nf_{s}\right)\right]^{2}}.\label{eq:35}\end{equation}

In order to relate the The PSD $S_{z}(f)$ of the averaged discrete
time series $z_{k}$ to the PSD of the sampled signal $y_{k}$ we
compute the Fourier Transform ${M^{*}\left(\mathcal{F}\right)}$ of
the digital filter ${m_{k}}$ where $\mathcal{F}$ is a normalized
frequency for the discrete time signals: $\mathcal{F}=f\cdot T$.
The impulse response of this filter is ${m_{k}=\frac{1}{m}\pi_{{m}}\left(k\right)}$,
where ${\pi_{{m}}\left(k\right)}$, is a discrete rectangular window
of length $m$ with all its coefficients equal to 1. This impulse
response is obtained from ${m\left(t,\tau\right)}$ by sampling it
with a sampling period $T$. The Fourier Transform ${M^{*}\left(\mathcal{F}\right)}$
is then: \begin{eqnarray}
M^{*}\left(\mathcal{F}\right) & = & \frac{1}{m}\sum_{k=0}^{m-1}{e^{{-2{i{\pi}}{k\mathcal{F}}}}}=\frac{1}{m}\frac{1-\exp\left(-2{i{\pi}}{\mathcal{F}m}\right)}{1-\exp\left(-2{i{\pi}\mathcal{F}}\right)}\nonumber \\
 & = & \frac{1}{m}\frac{\sin\left(\pi{m\mathcal{F}}\right)}{\sin\left({{\pi}\mathcal{F}}\right)}\exp\left[-i\pi\mathcal{F}\left(m-1\right)\right].\label{eq:defMf}\end{eqnarray}

We can notice that the frequency response of the discrete moving average
filter of equation (\ref{eq:defMf}) is different from that of the
continuous moving average filter of equation (\ref{eq:mov-ave}) when
replacing $\tau$ by $mT$.

As for the continuous time signals, this $M$ filter is not sufficient
to ensure the convergence of the variance for power law spectrum signals
with $\alpha<0$. Therefore, we introduce a digital version of the
continuous $D$ filter by choosing an impulse response $d_{k}$ as:
\begin{equation}
d_{k}=\left({\delta}_{k}-{\delta}_{{k-m}}\right)\label{eq:def_dk}\end{equation}
 where $\delta_{k}$ is a Dirac impulse of unity amplitude.

{As for the discrete time filter } ${m_{k}}${, the discrete filter
} ${d_{k}}$ {is obtained by sampling } ${d\left(t,\tau\right)}${
of equation (\ref{eq:defdttau}).}

The frequency response ${D^{{*}}\left(\mathcal{F}\right)}$ of the
digital filter ${d_{k}}$ is identical to that of the continuous filter
$D(\mathcal{F})$: \begin{equation}
D^{{*}}\left(\mathcal{F}\right)=1-e^{{-2{i{\pi}}{\mathcal{F}m}}}=2i\sin\left(\pi{\mathcal{F}m}\right)e^{{-{i{\pi}}{\mathcal{F}m}}}.\label{eq:defDsp}\end{equation}

When using $n$ difference filters we get the digital filter $d_{k}^{(n)}$
by sampling the continuous time filter $d^{{\left(n\right)}}\left(t,\tau\right)$
of equation (\ref{eq:defdnttau}): \begin{equation}
d_{k}^{{(n)}}=\sum_{p=0}^{n}(-1)^{k}C_{n}^{k}\delta_{k-pm}.\label{eq:39}\end{equation}

This impulse response could be obtained also by a digital convolution
(denoted by $\otimes$ in the following) of the filter ${d_{k}}$
in equation (\ref{eq:def_dk}) with itself $n$ times. The frequency
response $D^{*(n)}(\mathcal{F})$ of the filter $d_{k}^{(n)}$ is,
according to (\ref{eq:defDsp}), given by: \begin{equation}
D^{*(n)}(\mathcal{F})=\left[D^{{*}}\left(\mathcal{F}\right)\right]^{n}=2^{n}i^{n}\sin^{n}\left(\pi\mathcal{F}m\right)e^{-in\pi\mathcal{F}m}.\label{eq:40}\end{equation}

If we use the same normalization factor $c_{n}$ as the ones of equation
(\ref{eq:coefcn}), the output $u_{k}$ of the normalized filter $d_{k}^{(n)}/c_{n}$
is given by: \begin{equation}
u_{k}=\frac{1}{c_{n}}\left(d_{k}^{(n)}\otimes z_{k}\right)=\frac{1}{c_{n}}\left(d_{k}^{{(n)}}\otimes m_{k}\right)\otimes y_{k}.\label{eq:41}\end{equation}

{According to equation}{s (\ref{eq:41}), (\ref{eq:sigy_sy}),
(\ref{eq:defMf}) and (\ref{eq:40}), the true variance } ${{\sigma}_{{u}}^{2}\left(m\right)}$
{of the discrete signal } ${u_{k}}${ is related to the PSD }
${S_{y}\left(f\right)}${ of the discrete signal } ${y_{k}}${
by:} \begin{equation}
{\sigma}_{{u}}^{2}\left(m\right)=\frac{2^{{2n}}}{c_{{n}}^{2}m^{2}}\int_{-1/2}^{+1/2}\frac{\sin^{{2n+2}}\left(\pi{fm}\right)}{\sin^{2}\left({{\pi}f}\right)}S_{y}\left(f\right)df.\label{eq:42}\end{equation}

.

Equation (\ref{eq:42}) defines a stability true variance of discrete-time
signals in the general case. Percival proposed in \cite{percival2}
an identical formula to that obtained in (\ref{eq:42}) when $n=1$
in the case of Allan variance.

Comparing this expression to equation (\ref{eq:varUspec}) we can
notice that the denominator in (\ref{eq:42}) is ${m^{2}\sin\left({{\pi}f}\right)}$
while that of equation (\ref{eq:varUspec}) is ${\left(\pi\tau f\right)^{2}}$.
We have shown in equation (\ref{eq:PSDneg}) that ${S_{y}(f)\approx S_{Y}(f)/T}$.
This difference bewteen equations (\ref{eq:varUspec}) and (\ref{eq:42})
may let us think that the true variance ${{\sigma}_{{u}}^{2}\left(m\right)}${
of } ${u_{k}}$ is different from the variance ${{\sigma}_{{U}}^{2}\left(\tau\right)_{{(n)}}}$
of the continuous signal ${U^{{(n)}}\left(t,\tau\right)}$. Appendix
\ref{sect:AppxA} show a mathematical demonstration of the equivalence
of the discrete-time variance and the continuous-time variance.

{The above discret-time variance can be written versus the PSD of
the discrete-time error samples $x_{k}$. Using equation (\ref{eq:def_zk})
and (\ref{eq:30}) we can write:} \begin{equation}
{mz}_{k}=x_{{k}}-x_{k-m}=d_{k}^{{(1)}}{\otimes}x_{k}.\label{eq:47}\end{equation}

{Using this expression in (\ref{eq:41}) we can express } ${u_{k}}${
in terms of the phase measurement } ${x_{k}}${ under the simple
form:} \begin{equation}
u_{k}=\frac{1}{m\; c_{{n}}}\left(d_{k}^{{(n+1)}}{\otimes}x_{k}\right).\label{eq:48}\end{equation}

It's clear that equation (\ref{eq:48}) is simpler than equation (\ref{eq:41})
in terms of computation complexity because the filter ${d_{k}^{{(n)}}{\otimes}m_{k}}$
of equation (\ref{eq:41}) must be computed for each $m$ value while
the coefficients of the filter ${d_{k}^{{(n+1)}}}$ of equation (\ref{eq:48})
do not depend on the averaging factor $m$.

{According to equations (\ref{eq:48}), (\ref{eq:sigy_sy}) and (\ref{eq:40}),
the true variance } ${{\sigma}_{{u}}^{2}\left(m\right)}$ {of the
discrete signal } ${u_{k}}${ is related to the PSD } ${S_{{x}}\left(f\right)}${
of the discrete signal } ${x_{k}}${ by:} \begin{equation}
{\sigma}_{{u}}^{2}\left(m\right)=\frac{2^{{2n+2}}}{c_{{n}}^{2}m^{2}}\int_{-1/2}^{+1/2}{\sin^{{2n+2}}\left(\pi{fm}\right)S_{{x}}\left(f\right)}df.\label{eq:49}\end{equation}

This equation shows that the transition from the stability variance
of the continuous-time signal $X(t)$ given by equation (\ref{eq:22})
to the stability variance of discrete-time signal $x_{k}$ is done
very simply.

\subsection{Estimation of the Stability Variances of the Discrete-Time Signals}

{In order to estimate the variances presented in the last section
we use the sample variance of the zero mean discrete signal } ${u_{k}}${:}
\begin{equation}
{\hat{{\sigma}}}_{{u}}^{2}\left(m\right)=\frac{1}{N}\sum_{k=0}^{N-1}{|u_{k}|^{2}}\label{eq:50}\end{equation}
 where $N$ is the length of the time series ${u_{k}}$.

When the signal $u_{k}$ is obtained by filtering a signal of length
$L$ using a filter of length $p$, we must consider in (\ref{eq:50})
only $N=L-p+1$ unambiguous samples of $u_{k}$.

{Let } ${U_{k}}${ be the Discrete Fourier Transform (DFT) of the
discrete signal } ${u_{k}}${ defined by:} \begin{equation}
U_{{n}}=\sum_{k=0}^{N-1}{u_{k}e^{{-2{i{\pi}}\frac{{kn}}{N}}}}\;,\; n\in\left\{ 0,\cdots,N-1\right\} .\label{eq:51}\end{equation}

{The sample variance can be related to the }{DFT series using the
discrete Parseval's theorem:} \begin{equation}
\sum_{k=0}^{N-1}{|u_{k}|^{2}}=\frac{1}{N}\sum_{k=0}^{N-1}{|U_{k}|^{2}}.\label{eq:52}\end{equation}

The $U_{k}$ coefficients for $N/2<k<N$ represent the negative frequencies.
In the case of a real signal $u_{k}$, the coefficients $U_{k}$ are
symmetrical around $P=\left[\frac{N-1}{2}\right]$. We define a {}``one-sided''
set of DFT coefficients $\tilde{U}_{k}$ by: \begin{equation}
\tilde{U}_{k}=\left\{ \begin{array}{l}
\tilde{U}_{0}={\displaystyle \frac{U_{0}}{\sqrt{2}}}\\
\tilde{U}_{k}=U_{k},\qquad0\le k\le P-1\\
\tilde{U}_{P}=\left\{ \begin{array}{lcc}
U_{P} & \textrm{if} & N\quad\textrm{is odd}\\
{\displaystyle \frac{U_{P}}{\sqrt{2}}} & \textrm{if} & N\quad\textrm{is even.}\end{array}\right.\end{array}\right.\label{eq:53}\end{equation}

{The Parseval's }{theorem could be written then:} \begin{equation}
\sum_{k=0}^{N-1}|u_{k}|^{2}=\frac{2}{N}\sum_{k=0}^{P}\left|{\tilde{{U}}}_{k}\right|^{2}.\label{eq:54}\end{equation}

{According to equations (\ref{eq:41}) and (\ref{eq:47}), the DFT
coefficients } ${U_{{k}}}$ {of the time series } ${u_{k}}${
are related to that of } ${x_{k}}${ and } ${y_{k}}${ by:} \begin{equation}
U_{{k}}=\frac{1}{c_{{n}}}M^{*}\left(\frac{k}{N}\right)D^{{*\left(n\right)}}\left(\frac{k}{N}\right)Y_{{k}}=\frac{1}{m\; c_{{n}}}D^{{*\left(n+1\right)}}\left(\frac{k}{N}\right)X_{{k}}.\label{eq:55}\end{equation}

{The transition from equation (\ref{eq:41}) to the first part of
the above equation is valid under the assumption that discrete-time
signals are N-periodic. This means that the sample variance in the
frequency domain is equivalent to the sample variance in the time-domain
applied to an extended version (by periodization) of the discrete-time
signal. The first part of above equality gives when using (\ref{eq:defMf}),
(\ref{eq:40}), (\ref{eq:54}) and (\ref{eq:50}):} \begin{equation}
\hat{\sigma}_{F,u}^{2}(m)_{(n)}=\frac{2^{2n+1}}{c_{n}^{2}m^{2}N^{2}}\sum_{k=0}^{P}\frac{\sin^{2n+2}\left(\frac{\pi km}{N}\right)}{\sin^{2}\left(\frac{\pi k}{N}\right)}\left|\tilde{Y}_{k}\right|^{2}.\label{eq:56}\end{equation}

{To our knowledge, this is the first time that a relation between
the }{sample variance estimator of the frequency stability and the
DFT of discrete time series } ${y_{k}}$ {is established. It's worth
recalling that this equation is not a direct approximation to compute
the generic variance expression of equation (\ref{eq:varUspec}) by
discretization in the frequency domain as was proposed in \cite{Chang}
but it is the variance, according to the Parseval's theorem (\ref{eq:54}),
of a signal ${y_{k}}$ filtered in the frequency domain .}

Some works \cite{these_vernotte} have shown that using the numerical
integration in (\ref{eq:varUspec}) to estimate the Allan variance
($n=1$) leads to a biased estimator regarding the classical Allan
variance sample estimator. We will show at the end of this paper that
our formula (\ref{eq:56}) gives results which are nearly identical
to the classical sample estimators.

In fact, if we can consider that $Y(t)$ is band-limited to $f_{\textrm{max}}=\frac{1}{2T}$
then we can approximate the integral in equation (\ref{eq:varUspec})
in the Riemann sense by replacing the integration by the sum of the
surfaces of rectangles of width $\frac{1}{NT}$ at discrete frequencies
$f_{k}=\frac{k}{NT}$: \begin{equation}
\hat{\sigma}_{U}^{2}(mT)_{(n)}=\frac{2^{2n}NT}{c_{n}^{2}\pi^{2}m^{2}T^{2}}\sum_{k=0}^{P}\frac{\sin^{2n+2}\left(\frac{\pi km}{N}\right)}{k^{2}}\hat{S}_{Y}\left(\frac{k}{NT}\right).\label{eq:57}\end{equation}

where $\hat{S}_{Y}(f)$ is an estimator of the PSD $S_{Y}(f)$. If
we use ${2T|{\tilde{Y}}_{k}|^{2}/N}${ as an estimator of } ${S_{Y}\left(f_{k}\right)}${
then equation (\ref{eq:57}) becomes :

\begin{equation}
\hat{\sigma}_{U}^{2}(mT)_{(n)}=\frac{2^{2n+1}}{c_{n}^{2}\pi^{2}m^{2}}\sum_{k=0}^{P}\frac{\sin^{2n+2}\left(\frac{\pi km}{N}\right)}{k^{2}}|{\tilde{Y}}_{k}|^{2}.\label{eq:57a}\end{equation}
It's clear that equations (\ref{eq:57a}) and (\ref{eq:56}) are different.
This difference could by explained by the fact that equation (\ref{eq:varUspec})
is given versus } ${S_{Y}\left(f\right)}${ which is not observable
directly while equation (\ref{eq:57}) use } ${|{\tilde{Y}}_{k}|^{2}}${,
an estimator of the PSD of the averaged and sampled version of $Y(t)$.
In other words, averaging according to equation (\ref{eq:yk_xt})
is considered when using } ${|{\tilde{Y}}_{k}|^{2}}${ in equation
(\ref{eq:56}) while } ${S_{Y}\left(f\right)}${ in equations (\ref{eq:varUspec})
and (\ref{eq:57}) is considered before averaging according to equation
(\ref{eq:4}).}

{In order to relate equation (\ref{eq:57}) to equation (\ref{eq:varUspec})
we suppose that } ${S_{Y}\left(f\right)}${ is band limited. In
this case, there is no aliasing in equation (\ref{eq:PSDper}) and
it could be written:} \begin{equation}
S_{y}\left(f\right)=\frac{1}{T}S_{Y}\left(f\right)\frac{\sin^{2}\left(\pi{Tf}\right)}{\left(\pi{Tf}\right)^{2}}\qquad\textrm{for}\quad0\le f\le\frac{1}{2T}.\label{eq:58}\end{equation}

The DFT coefficients ${\tilde{Y}_{k}}$ could be considered as an
estimator of the PSD ${S_{y}\left(f\right)}$ of the discrete signal
${y_{k}}$ at discrete frequencies ${f_{k}}$: \begin{equation}
{{\hat{{S}}}_{y}\left(f_{k}\right)=\frac{2}{N}|{\tilde{Y}}_{k}|^{2}}\qquad\textrm{for}\quad0\le k\le P.\label{eq:59}\end{equation}

This equation is known in the literature as the periodogram spectrum
estimator. The factor 2 in (\ref{eq:59}) is due to the fact that
the PSD $S_{y}(f)$ is one-sided.

Replacing equations (\ref{eq:59}) in (\ref{eq:58}), we get an estimator
of the PSD $S_{Y}(f)$ of the band-limited continuous time signal
$Y(t)$: \begin{equation}
\hat{S}_{Y}\left(\frac{k}{NT}\right)=\frac{2T\left(\pi k\right)^{2}}{N^{3}\sin^{2}\left(\frac{\pi k}{N}\right)}\left|\tilde{Y}_{k}\right|^{2}.\label{eq:60}\end{equation}

{Using this expression in}{ equation (\ref{eq:57}) leads to an
expression identical to the sample variance of equation (\ref{eq:56}).
This interesting result could be written as:} \begin{equation}
{\hat{{\sigma}}}_{{U}}^{2}\left({mT}\right)_{{\left(n\right)}}={\hat{{\sigma}}}_{{u}}^{2}(m).\label{eq:61}\end{equation}

{In other words, the sample variance of equation (\ref{eq:56}) is
equal to the integral of equation (\ref{eq:varUspec}) for a band-limited
\emph{Y(t) }when evaluated in the Riemann sense over the interval
} ${f\in\left[0,1/2T\right]}${ by using the periodogram of } ${y_{k}}$
{as an estimator of the PSD } ${S_{Y}\left(f\right)}${ of Y(t)
according to equation (\ref{eq:60}).}

The second part of equation (\ref{eq:55}) gives when using (\ref{eq:40}),
(\ref{eq:54}) and (\ref{eq:50}): \begin{equation}
\hat{\sigma}_{u}^{2}(m)=\frac{2^{2n+3}}{c_{n}^{2}m^{2}N^{2}}\sum_{k=0}^{P}\sin^{2n+2}\left(\frac{\pi km}{N}\right)\left|\tilde{X}_{k}\right|^{2}.\label{eq:62}\end{equation}

{When $X(t)$ is band-limited, equation (\ref{eq:62}) can be obtained
directly from equation (\ref{eq:22}) using a Riemann sum and replacing
} ${S_{{x}}\left(f\right)}${ by the periodogram of the discrete
signal } ${x_{k}}${.}

\begin{figure}
\centering \includegraphics[width=1\linewidth]{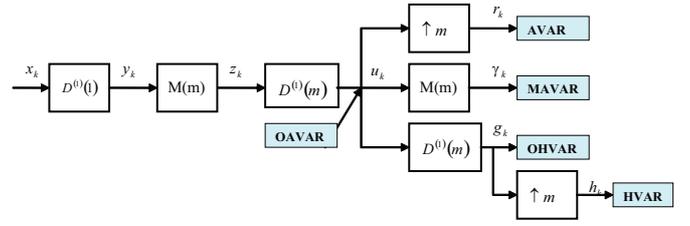} 

\caption{The processing chain of the stability variances, $M(m)$ is a moving
average filter of length $m$. $D^{(n)}(m)$ is a difference filter
of order $n$ and lag $m$. $\uparrow m$ is the decimation by a factor
$m$ operator.}

\label{fig:1} 
\end{figure}

{In the following of this paper we express the different stability
variances in the discrete time using the signal } ${u_{k}}${. Figure
\ref{fig:1} shows the different filters involved in the computation
of theses stability variances.}

\subsection{The Overlapping Allan Variance (OAVAR)}

\label{sect:OAVAR_f} {This is a special case of the above processing
when the order of the difference filter $n$ is equal to one. The
normalization factor } ${c_{{1}}}${ is given by equation (\ref{eq:coefcn})
and is equal to } ${\sqrt{2}}${. The filter } ${d_{k}^{{(2)}}}${
of equation (\ref{eq:48}) is equal to } ${{\delta}_{k}-2{\delta}_{{k+m}}+{\delta}_{{k+2m}}}${.
The signal } ${u_{k}}${ is given by:} \begin{equation}
u_{k}=\frac{1}{m\sqrt{2}}\left(x_{{k+2m}}-2x_{{k+m}}+x_{k}\right).\label{eq:63}\end{equation}

{Let $N$ the length of the discrete time series } ${y_{k}}${.
The } ${d_{k}^{{(2)}}}${ filter length is $2m$ and the output
} ${u_{k}}$ {length is $N-2m+1$.}

{According to equation (\ref{eq:50}), the sample variance of }
${u_{k}}${ is:}

\begin{eqnarray}
 &  & \textrm{OAVAR}(m)={\hat{{\sigma}}}_{{u}}^{2}\left(m\right)\label{eq:64}\\
 &  & =\frac{1}{2m^{2}\left(N-2m+1\right)}\sum_{k=0}^{N-2m}\left(x_{{k+2m}}-2x_{{k+m}}+x_{k}\right)^{2}\nonumber \end{eqnarray}

which is the classical estimator of the Overlapping estimator of Allan
Variance \cite{howe2}.

{The computation in (\ref{eq:64}) from } ${x_{k}}${ requires
four additions and one multiplication for each term inside the sum.
The sum over $k$ requires $N-2m+1$ addition. The whole computation
requires roughly $5\times N$ operation and is linear in $N$.}

{When the available measurement are frequency fluctuations (} $\left|y_{k}\right|${),
it's more efficient (in number of floating point operations but not
in memory use) to compute the phase signal } ${x_{k}}${ using (\ref{eq:31})
and then use (\ref{eq:64}) to compute the OAVAR variance than to
compute } ${z_{k}}${ from } ${y_{k}}${ and then } ${u_{k}}${.}

{Replacing $n$ by 1 in equation (\ref{eq:56}) we get an expression
of the Overlapped Allan Variance versus the one-sided set of DFT coefficient
} ${{\tilde{Y}}_{k}}$ {of the measurement time series } ${y_{k}}${
by:} \begin{equation}
\textrm{OAVAR}(m)_{F}={\hat{{\sigma}}}_{{F,u}}^{2}\left(m\right)=\frac{4}{m^{2}N^{2}}\sum_{k=0}^{P}\frac{\sin^{{4}}\left(\frac{\pi{km}}{N}\right)}{\sin^{2}\left(\frac{{{\pi}k}}{N}\right)}|{\tilde{Y}}_{k}|^{2}.\label{eq:65}\end{equation}

{The }{DFT computation complexity is N log(N) when using a Fast
Fourier Transform (FFT) algorithm. But the most CPU consuming in (\ref{eq:65})
is the computation of the sine trigonometric functions inside the
sum symbol. It's trivial that the computation using equation (\ref{eq:64})
is more efficient than using equation (\ref{eq:65}).}

{It's worth recalling that the discrete time formula (\ref{eq:64})
use $N-2m+1$ terms. The largest acceptable $m$ value is $N/2$.
In this case the variance is estimated from one sample only. The DFT
formula (\ref{eq:65}) use $P$ terms whatever the $m$ value. When
$m=N/2$ the half of the sine terms in (\ref{eq:65}) is null. The
computation of the confidence levels when using equation (\ref{eq:65})
has shown that the confidence levels are better than that of the discrete
time formula of equation (\ref{eq:64}) because filtering in frequency
domain use all the available samples while filtering in the time domain
use $N$ minus the filter length samples. In fact, Filtering in the
DFT domain is done by multiplication of the DFT. This multiplication
is equivalent to circular convolution in the time domain. Circular
or cyclic convolution of two signal of length $N$ is equivalent to
classical sum convolution with indices modulo $N$. This means that
DFT formula (\ref{eq:65}) is equivalent to a kind of Total Variance
\cite{totvar} where the series } ${y_{k}}${ is extended by periodic
(circular) repetitions. The Total Hadamard Variance \cite{howe} uses
an extended version of } ${y_{k}}${ where the extension use a reflected
copy of } ${y_{k}}${.}

\subsection{The {}``Non Overlapping'' Allan Variance (AVAR)}

{The {{}``}Non overlapping{\textquotedblright} Allan variance
is a special case of the classical Allan Variance that doesn't use
overlapped values when computing } ${\sum{u_{k}^{2}}}${ in the
sample variance of } ${u_{k}}${. This means that only $(N-2m+1)/m$
values are considered when forming the sum.}

{In other words, the Allan Variance AVAR is obtained from } ${u_{k}}${
by a decimation operation of order $m$ (See Figure \ref{fig:1}).
If we start the decimation at $k=0$ we can use $N/m-1$ values. The
decimated signal } ${r_{k}}${ is given by:} \begin{equation}
{r_{k}=u_{{km}}\;,\;0\le k\le\frac{N}{m}-2}.\label{eq:66}\end{equation}

{Replacing (\ref{eq:66}) in (\ref{eq:64}) we get the non overlapping
Allan variance as the sample variance of } ${r_{k}}${:} \begin{eqnarray}
 & \textrm{AVAR}(m)=\hat{\sigma}_{r}^{2}(m)\label{eq:67}\\
 & =\frac{1}{2m(N-m)}\sum_{k=0}^{N/m-2}\left(x_{(k+2)m}-2x_{(k+1)m}+x_{km}\right)^{2}\nonumber \end{eqnarray}

{It's obvious that the AVAR requires }{less computation than the
OAVAR. In fact, for each $m$ value there are $N/m-1$ terms. The
largest acceptable $m$ value is $N/2$. In this case the sample variance
is estimated from one sample only. The confidence levels for AVAR
and OAVAR are equals for $m=1$ and $m=N/2$. Values of m between
$m=1$ and $m=N/2$ give a better confidence levels in the OAVAR than
in the AVAR variance.}

{ Because the OAVAR confidence levels are globally better than those
of AVAR , the only interest to use the AVAR instead of OAVAR is its
computation efficiency.}

{Though decimation operation of equation (\ref{eq:66}) is very simple
in the time domain it has no interest in the frequency domain. In
fact, the computation of the DFT coefficient } ${R_{k}}$ {of }
${r_{k}}${ versus the DFT coefficients of } ${u_{k}}${ is given
by:} \begin{equation}
R_{n}=\frac{1}{m}\sum_{k=0}^{m-1}U_{\left(n-pN/m\right)}\quad\textrm{Modulo }\quad N.\label{eq:68}\end{equation}

{Computing the sample variance of } ${r_{k}}${ according to (\ref{eq:54})
require an additional loop to compute } ${R_{k}}${. For this reason
we don't propose a formula to compute the AVAR in the frequency domain
as we did for OAVAR in equation (\ref{eq:65}).}

\subsection{The Modified Allan Variance (MAVAR)}

The modified Allan Variance was introduced \cite{modallan} to overcome
the relatively poor discrimination capability of the Allan variance
against white and flicker phase noise.

Let ${{\gamma}_{k}}$ be the signal obtained from ${u_{k}}$ by a
moving average filter $M(m)$ of length $m$ (See Figure \ref{fig:1}):
\begin{equation}
{\gamma}_{k}=\frac{1}{m}\left(u_{k}+u_{{k+1}}+\cdots+u_{{k+m-1}}\right).\label{eq:69}\end{equation}

{Using } ${u_{k}}${ expression from (\ref{eq:63}) in (\ref{eq:69})
we get:} \begin{eqnarray}
\sqrt{2}m^{2}\gamma_{k} & = & x_{k}+\cdots+x_{{k+m-1}}\nonumber \\
 &  & -2\left(x_{{k+m}}+\cdots+x_{{k+2m-1}}\right)\nonumber \\
 &  & +x_{{k+2m}}+\cdots+x_{{k+3m-1}}.\label{eq:70}\end{eqnarray}

{Using this expression directly to compute } ${{\gamma}_{k}}${
requires a summation loop with $3*(m+1)$ floating point operation.
The biggest acceptable $m$ value in this equation is $m=N/3$. This
yields a computation complexity of } ${N^{2}}${.}

{In order to reduce the computation complexity we propose a recursive
formula. Expressing } ${A_{{k+1}}=\sqrt{2}m^{2}{\gamma}_{{k+1}}}${
using equation (\ref{eq:70}) we can write:} \begin{equation}
A_{{k+1}}=A_{k}+x_{{k+3m}}-3x_{{k+2m}}+3x_{{k+m}}-x_{k}\label{eq:71}\end{equation}
 with a starting $A_{0}$ value computed using (\ref{eq:70}) with
$k=0$.

Allan \cite{allan_eftf} already proposed a recursive method in order
to reduce the computation complexity of the Modified Allan Variance
without giving the details of the recursive equation. 

{The computation complexity of } ${A_{k}}${ according to (\ref{eq:71})
is linear in $N$.}

{The length of the time series } ${u_{k}}${ is $N-2m+1$ and the
length of the filter $M(m)$ is $m$. we conclude that the length
of } ${{\gamma}_{k}}${ is $N-3m+2$.}

{The Modified Allan Variance MAVAR is }{the sample variance of
} ${{\gamma}_{k}}${:} \begin{equation}
\textrm{MAVAR}(m)={\hat{{\sigma}}}_{{\gamma}}^{2}\left(m\right)=\frac{1}{2m^{{4}}\left(N-3m+2\right)}\sum_{k=0}^{N-3m-1}|A_{k}|^{2}.\label{eq:72}\end{equation}

{The PSD } ${S_{{\gamma}}\left(f\right)}${ of the discrete time
signal } ${{\gamma}_{k}}${ is related to that of } ${y_{k}}${
by:} \begin{eqnarray}
S_{{\gamma}}\left(f\right) & = & |M^{*}\left(f\right)D^{{*}}\left(f\right)M^{*}\left(f\right)|^{2}S_{y}\left(f\right)\nonumber \\
 & = & \frac{2\sin^{{6}}\left(\pi{mf}\right)}{m^{{4}}\sin^{{4}}\left({{\pi}f}\right)}S_{y}\left(f\right).\label{eq:73}\end{eqnarray}

{The DFT coefficient}{s } ${{\Gamma}_{{n}}}$ {of the series
} ${{\gamma}_{k}}${ are given by:} \begin{equation}
{\Gamma}_{{n}}=M^{*}\left(\frac{n}{N}\right)D^{{*}}\left(\frac{n}{N}\right)M^{*}\left(\frac{n}{N}\right)Y_{{n}}=\frac{\sqrt{2}\sin^{{3}}\left({{\pi}m}\frac{n}{N}\right)}{m^{2}\sin^{2}\left(\pi\frac{n}{N}\right)}Y_{{n}}.\label{eq:74}\end{equation}

{According to the Parseval}{'s equation (\ref{eq:54}) for the
series } ${{\gamma}_{k}}${ we can express the MAVAR versus the
one-sided set of DFT coefficients of the measured signal by:} \begin{equation}
\textrm{MAVAR}(m)_{F}={\hat{{\sigma}}}_{{F,\gamma}}^{2}\left(m\right)=\frac{4}{m^{{4}}N^{2}}\sum_{k=0}^{P}\frac{\sin^{{6}}\left(\frac{\pi{mk}}{N}\right)}{\sin^{{4}}\left(\frac{{{\pi}k}}{N}\right)}|{\tilde{Y}}_{k}|^{2}.\label{eq:75}\end{equation}

{As for the OAVAR formula in the frequency domain this equation is
an estimator of the MAVAR in the frequency domain. The main difference
with the discrete time formula (\ref{eq:72}) is the number of terms
involved in the sum: $P$ in the case of equation (\ref{eq:75}) and
$N-2m+1$ in equation (\ref{eq:72}).}

\subsection{The Overlapping Hadamard Variance (OHVAR)}

{This is a special case of the above processing when the order of
the difference filter $n$ is equal to two. The normalization factor
} ${c_{{2}}}${ is given by equation (\ref{eq:coefcn}) and is equal
to } ${\sqrt{6}}${. The filter } ${d_{k}^{{(3)}}}${ of equation
(\ref{eq:48}) is equal to } ${{\delta}_{k}-3{\delta}_{{k+m}}+3{\delta}_{{k+2m}}-{\delta}_{{k+4m}}}${.
We denote } ${g_{k}=u_{k}}${ where } ${u_{k}}${ is given by
(\ref{eq:48}) with $n=2$:}

\begin{equation}
{g_{k}=\frac{1}{\sqrt{6}m}\left(x_{{k+3m}}-3x_{{k+2m}}+3x_{{k+m}}-x_{k}\right)}.\label{eq:76}\end{equation}

{Let $N$ the length of the discrete time series } ${y_{k}}${.
The } ${d_{k}^{{(3)}}}${ filter length is 3m and the output }
${g_{k}}$ {length is $N-3m+1$.}

{The Overlapping Hadamard Variance is the sample variance of } ${g_{k}}${:}
\begin{eqnarray}
\textrm{OHVAR}(m) & = & {\hat{{\sigma}}}_{{g}}^{2}\left(m\right)\\
 & = & \frac{1}{6m^{2}\left(N-3m+1\right)}\nonumber \\
 &  & \times\sum_{k=0}^{N-3m}{\left(x_{{k+3m}}-3x_{{k+2m}}+3x_{{k+m}}-x_{k}\right)^{2}}.\nonumber \end{eqnarray}

Replacing $n$ by 2 in equation (\ref{eq:56}) we get an expression
of the Overlapping Hadamard Variance versus the one-sided set of DFT
coefficient $\tilde{Y}_{k}$ of the measurement time series ${y_{k}}$
by: \begin{equation}
\textrm{OHVAR}_{F}\left(m\right)={\hat{{\sigma}}}_{{F,g}}^{2}\left(m\right)=\frac{16}{3m^{2}N^{2}}\sum_{k=0}^{P}{\frac{\sin^{{6}}\left(\frac{\pi{mk}}{N}\right)}{\sin^{2}\left(\frac{{{\pi}k}}{N}\right)}|{\tilde{Y}}_{k}|^{2}}.\label{eq:78}\end{equation}

{As for formulas (\ref{eq:65}) and (\ref{eq:75}), equation (\ref{eq:78})
is a new formula that allows to compute the OHAVAR in the frequency
domain.}

{It's clear from equation (\ref{eq:76}) that the Hadamard variance
estimator in the time domain cancels linear drifts. In fact, if }
${y_{k}=k}${ then } ${x_{k}=k\left(k-1\right)/2}${ according
to equation (\ref{eq:31}). Replacing this value in equation (\ref{eq:76})
leads to } ${g_{k}=0}${ whatever the value of m.}

\subsection{The Hadamard Variance (HVAR)}

{The Hadamard variance is a special case of the Overlapping Hadamard
Variance that doesn't use overlapped values when computing } ${\sum{g_{k}^{2}}}${
in the sample variance of } ${g_{k}}${. This means that only $(N-3m+1)/m$
values are considered when forming the sum.}

{In other words, the Hadamard Variance HVAR is obtained from } ${g_{k}}${
by a decimation operation of order $m$ (See Figure \ref{fig:1}).
If we start the decimation at $k=0$ we can use $N/m-2$ values. The
decimated signal } ${h_{k}}${ is given by:} \begin{equation}
{h_{k}=g_{{km}}\;,\;0\le k\le\frac{N}{m}-3}.\label{eq:79}\end{equation}

{Replacing (\ref{eq:79}) in (\ref{eq:76}) we get the non overlapping
Hadamard variance as the sample variance of } ${h_{k}}${:} \begin{eqnarray}
\textrm{HVAR}(m) & = & {\hat{{\sigma}}}_{{h}}^{2}\left(m\right)\nonumber \\
 & = & \frac{1}{6m\left(N-2m\right)}\sum_{k=0}^{N/m-3}\left(x_{{(k+3)m}}\right.\nonumber \\
 &  & \left.-3x_{{(k+2)m}}+3x_{{(k+1)m}}-x_{{km}}\right)^{2}.\label{eq:80}\end{eqnarray}

{As for the Non }{Overlapping Allan Variance AVAR we don't propose
a formula in the frequency domain for HVAR because the decimation
operation doesn't simplify computation in the frequency domain as
it does in the time-domain.}

\section{Frequency variances Equivalent Degree of Freedom}

We can express the frequency-domain variance estimator by the general
form :

\begin{equation}
\Psi=\sum_{k=0}^{P}{H_{k}(n,m)\frac{{\left|\tilde{Y}\right|^{2}}}{N}}\label{eq:B1}\end{equation}

Where $n$ is the difference filter order, $m$ is the averaging factor
and $H_{k}(n,m)$ is given by :

\begin{equation}
H_{k}(n,m)=\frac{2^{2n+1}}{c_{n}^{2}m^{2}N}\frac{\sin^{2n+2}\left(\frac{\pi km}{N}\right)}{\sin^{2}\left(\frac{\pi k}{N}\right)}\label{eq:B2}\end{equation}

for the non-modified variances and :

\begin{equation}
H_{k}(n,m)=\frac{2^{2n+1}}{c_{n}^{2}m^{4}N}\frac{\sin^{2n+4}\left(\frac{\pi km}{N}\right)}{\sin^{4}\left(\frac{\pi k}{N}\right)}\label{eq:B3}\end{equation}

for the modified variances.

The quantity $\left|\tilde{Y}\right|^{2}/N$ is the periodogram $P(f)$
evaluated at discrete frequency values $f_{k}=$$\frac{k}{N}$. Equation
( \ref{eq:B1} ) can be written as :

\begin{equation}
\Psi=\sum_{k=0}^{P}{H_{k}(n,m)P\left(f_{k}\right)}\label{eq:B4}\end{equation}

The periodogram $P(f)$ ~is an estimator of the PSD $S_{y}(f)$ :
$P(f)=\hat{S}\left(f\right)$ .

We estimate the Equivalent Degree of Freedom (edf) of $\Psi$ by :

\begin{equation}
edf=\frac{2\left(E\left(\Psi\right)\right)^{2}}{Var\left(\Psi\right)}\label{eq:B5}\end{equation}

The mean value $E\left(\Psi\right)$ is given by :

\begin{equation}
E\left(\Psi\right)=\sum_{k=0}^{P}{H_{k}(n,m)E\left(P\left(f_{k}\right)\right)}\label{eq:B6}\end{equation}

It is well know that the periodogram is a biased estimator of the
PSD $S_{y}(f)$ and that :

\begin{equation}
E\left(P\left(f\right)\right)=W_{B}\left(f\right)\otimes S\left(f\right)\label{eq:B7}\end{equation}

Where $W_{B}\left(f\right)$ is the Bartlett window defined by :

\begin{equation}
W_{B}\left(f\right)=\frac{\sin^{2}\left(\pi Nf\right)}{N\sin^{2}\left(\pi f\right)}\label{eq:B8}\end{equation}

and $\otimes$ denotes the circular convolution defined by :

\begin{equation}
W_{B}\left(f\right)\otimes S\left(f\right)=\int_{-1/2}^{1/2}W_{B}\left(\theta\right)S\left(f-\theta\right)d\theta\label{eq:B9}\end{equation}

It\textquoteright{}s clear that the periodogram is asymptotically
unbiased since as $N$ becomes very large $W_{B}\left(f\right)$ approaches
an impulse in the frequency domain. Then we can write for large \emph{N}
:

\begin{equation}
E\left(\Psi\right)\cong\sum_{k=0}^{P}{H_{k}(n,m)S\left(f_{k}\right)}\label{eq:B10}\end{equation}

and for power law spectrum :

\begin{equation}
E\left(\Psi\right)=h_{\alpha}\sum_{k=0}^{P}{H_{k}(n,m)\left(\frac{k}{2N}\right)^{\alpha}}\label{eq:B11}\end{equation}

The variance $Var\left(\Psi\right)$ is given by :

\begin{equation}
\begin{array}{c}
Var\left(\Psi\right)=E\left(\Psi^{2}\right)\\
=\sum\limits _{k=0}^{P}\sum\limits _{j=0}^{P}{H_{k}(n,m)H_{j}(n,m)Cov\left(P\left(f_{k}\right),P\left(f_{j}\right)\right)}\end{array}\label{eq:B12}\end{equation}

The covariance of the periodogram is given by :\begin{eqnarray}
 & Cov\left(P\left(f_{1}\right),P\left(f_{2}\right)\right)=\label{eq:B13}\\
 & S_{y}\left(f_{1}\right)S_{y}\left(f_{2}\right)\left(\frac{\sin\left(\pi N\left(f_{1}-f_{2}\right)\right)}{N\sin\left(\pi\left(f_{1}-f_{2}\right)\right)}\right)^{2}\nonumber \end{eqnarray}

Replacing $f_{1}$ by $f_{k}=\frac{k}{N}$ and $f_{2}$ by $f_{j}=\frac{j}{N}$
in equation we get :

\begin{eqnarray}
 & Cov\left(P\left(f_{k}\right),P\left(f_{j}\right)\right)=\label{eq:B14}\\
 & S_{y}\left(f_{k}\right)S_{y}\left(f_{j}\right)\left(\frac{\sin^{2}\left(\pi\left(k-j\right)\right)}{N^{2}\sin^{2}\left(\frac{\pi}{N}\left(k-j\right)\right)}\right)\nonumber \end{eqnarray}
Therefore, the covariance (\ref{eq:B14}) is is seen to go to zero
when $k\neq j$ . The variance is therefore :

\begin{equation}
Var(\Psi)=\sum_{k=0}^{P}H_{k}^{2}(n,m)S_{y}^{2}(f_{k})\label{eq:B15}\end{equation}

The \emph{edf }is, according to (\ref{eq:B5}), given by :

\begin{equation}
edf=\frac{2(\sum_{k=0}^{P}H_{k}(n,m)S_{y}(f_{k}))^{2}}{\sum_{k=0}^{P}H_{k}^{2}(n,m)S_{y}^{2}(f_{k})}\label{eq:B16}\end{equation}

For power law spectrum we get : 

\begin{equation}
edf=\frac{2(\sum_{k=0}^{P}k^{\alpha}H_{k}(n,m))^{2}}{\sum_{k=0}^{P}k^{2\alpha}H_{k}^{2}(n,m)}\label{eq:B17}\end{equation}

With $H_{k}\left(n,m\right)$ given by (\ref{eq:B3}) for the modified
variances and (\ref{eq:B2}) for the non-modified variances.

\section{Time Domain versus Frequency Domain: Numerical Results}

{We have simulated time series }{data } ${y_{k}}${ of length
$N=400000$, $2000000$ and $65536$ for the different power law spectra
for } ${-4\le\alpha\le2}${. Table \ref{tab:4} and \ref{tab:5}
show the computation time on a personnal computer (pentium IV or equivalent
@ 2.8 GHz) in ms of the different stability variances mentioned in
this paper.} The computation time of the FFT was included in the
computation time of the frequency variances.

\begin{table}
\centering 

\caption{Computation time in ms of the different stability variances, $N=400000$}

\begin{tabular}{|c|c|c|c|c|c|}
\hline 
 & AVAR  & OAVAR  & MAVAR  & HVAR  & OHVAR \tabularnewline
\hline 
Time Domain  & 16  & 47  & 78  & 16  & 47 \tabularnewline
\hline 
Frequency Domain  & --  & 265  & 265  & --  & 265 \tabularnewline
\hline
\end{tabular}\label{tab:4} 
\end{table}

\begin{table}
\centering 

\caption{Computation time in ms of the different stability variances, $N=2000000$}

\begin{tabular}{|c|c|c|c|c|c|}
\hline 
 & AVAR  & OAVAR  & MAVAR  & HVAR  & OHVAR \tabularnewline
\hline 
Time Domain  & 63  & 265  & 484  & 63  & 360 \tabularnewline
\hline 
Frequency Domain  & --  & 1453  & 1500  & --  & 1485\tabularnewline
\hline
\end{tabular}\label{tab:5} 
\end{table}

{For the computation in the frequency domain we used the FFT algorithm
of Cooley and Tuckey\cite{fft}. The FFT computation time }{is 45
ms for $N=400000$ and 250 ms for $N=2000000$.}

We presented in equation (\ref{eq:56}) a new way to compute the different
stability variances using the DFT of the data. We demonstrated that
this equation is equivalent to the equations in the time domain with
a slight difference in the number of samples when computing the sample
variance. For example, equation (\ref{eq:64}) in the time domain
uses only unambiguous samples in the sense that a filter of length
$2m$ will produce $N-2m+1$ unambiguous output samples when applied
to an input data of length $N$. 

In the following we present numerical results of the different frequency
domain variances estimators presented in this paper. The error bars
on the plots were computed using one sigma Chi-squared $\chi^{2}$distribution
with an equivalent degree of freedom (\emph{edf}) estimated by making
Monte Carlo simulations of 1000 trials.

\subsection{OAVAR}

\begin{figure}
\centering \includegraphics[width=1\linewidth]{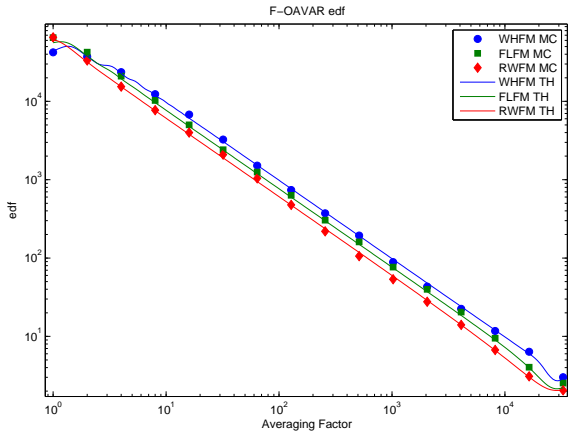} 

\caption{F-OAVAR \emph{edf} for three noise types for sequences of length $N=65536$.
WHFM for White frequency noise, FLFM for Flicker frequency noise and
RWFM for Random Walk frequency noise. The continous lines (denoted
{}``TH'' on the Figure legend) represent the theoretical \emph{edf}
computed by equation (\ref{eq:B17}). The symbols (denoted {}``MC''
on the Figure legend) represent the \emph{edf }obtained by Monte Carlo
simulation with 1000 trials.}

\label{fig:edf} 
\end{figure}

Figure (\ref{fig:edf}) depicts the \emph{edf }of the Overlapping
Allan Variance computed in the frequency domain (F-OAVAR) for three
noise types: a White frequency noise (WHFM), a Flicker frequency noise
(FLFM) and a Random Walk frequency noise (RWFM). It shows a very good
agreement between the theoretical \emph{edf }formula of equation (\ref{eq:B17})
and the \emph{edf }obtained by Monte Carlo simulations. 

\begin{figure}
\centering \includegraphics[width=1\linewidth]{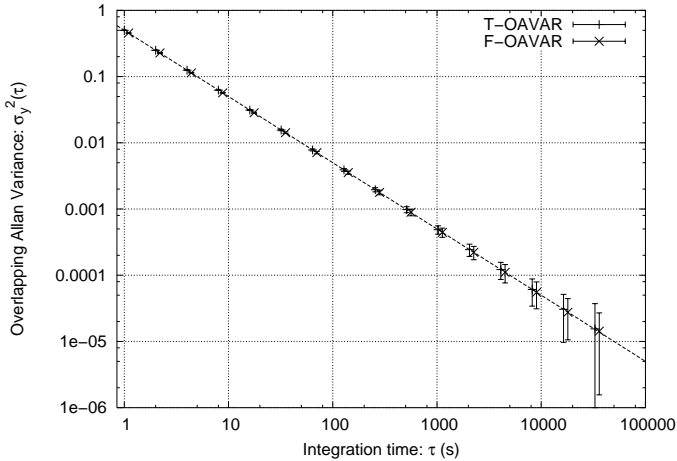} 

\caption{OAVAR computed in the time domain and in the frequency domain for
a White Frequency Noise sequence of length $N=65536$. The spectral
OAVAR estimates were slightly shifted in order to be distinguished
from the time OAVAR estimates. The dashed continuous line represents
the theoretical response $h_{0}/(2\tau)$.}

\label{fig:avar_ft_blc} 
\end{figure}

Figure (\ref{fig:avar_ft_blc}) compares the Overlapping Allan variance
of a white frequency noise sequence computed in the time domain and
in the frequency domain from relationship (\ref{eq:65}). No bias
is visible between these computations and the theoretical response
(less than 1 \%). On the other side, the error bars of OAVAR computed
in the frequency domain are clearly smaller as the ones of OAVAR computed
in the time domain, as expected in section \ref{sect:OAVAR_f}. Table
\ref{tab:deglib_blc} shows the equivalent degrees of freedom (\textit{edf})
of the Total Variance and the OAVAR estimates in the time domain (T-OAVAR)
and in the frequency domain (F-OAVAR), assuming a Chi-square statistics
\cite{lesage}. For the highest $\tau$ value ($\tau=N/2$), the\emph{
edf} of the spectral estimate is 3 times higher than the\emph{ edf
}of the time estimate, i.e. the spectral estimate is $\sqrt{3}$ times
more accurate than the time estimate.

\begin{table}
\centering 

\caption{Comparison of the equivalent degrees of freedom (\emph{edf}) of the
time T-OAVAR estimates, the spectral F-OAVAR estimates and the Total
variance estimates for a White Frequency Noise sequence of length
$N=65536$.}

\label{tab:deglib_blc} \begin{tabular}{|r|r|r|r|}
\hline 
$\mathbf{\tau}$  & T-OAVAR  & F-OAVAR  & TotVar\tabularnewline
\hline 
1 & 46591 & 42297 & 45368\tabularnewline
2 & 40640 & 37232 & 34379\tabularnewline
4 & 24186 & 23639 & 22460\tabularnewline
8 & 11870 & 12338 & 11451\tabularnewline
16 & 5865 & 6786 & 6375\tabularnewline
32 & 2937 & 3255 & 2945\tabularnewline
64 & 1493 & 1515 & 1555\tabularnewline
128 & 746 & 740 & 832\tabularnewline
256 & 383 & 372 & 414\tabularnewline
512 & 199 & 194 & 215\tabularnewline
1024 & 93 & 89 & 104\tabularnewline
2048 & 43 & 43 & 53\tabularnewline
4096 & 20 & 22 & 26\tabularnewline
8192 & 10 & 12 & 12\tabularnewline
16384 & 4 & 6.4 & 6.2\tabularnewline
32768 & 1.0 & 3.0 & 2.9\tabularnewline
\hline
\end{tabular}
\end{table}

Such an advantage is particularly useful for detecting and measuring
the level of the low frequency noises (e.g. random walk FM) sooner
as with time variances, i.e. for shorter duration. Considering that
the \emph{edf} decreases approximately as $\tau^{-1}$, an estimator
with an edf 3 times higher than another one provides a noise level
estimation $\sqrt{3}$ times sooner than the other one (e.g. 7 month
instead of 1 year) with the same accuracy.

\begin{figure}
\centering \includegraphics[width=1\linewidth]{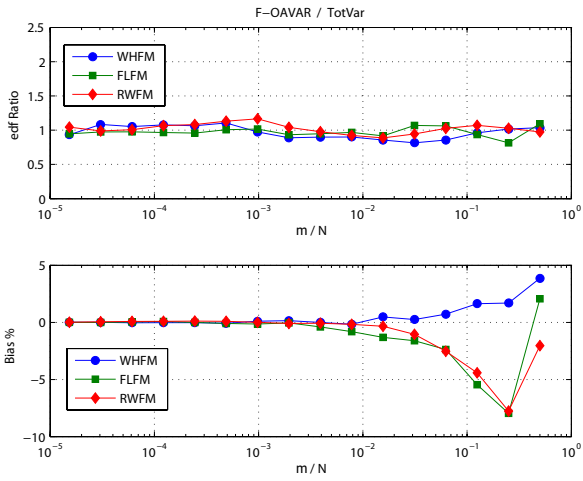} 

\caption{Comparaison of the F-OAVAR and the Total variance for three noise
types. The upper plot depicts the \emph{edf} ratio and the lower plot
depicts the bias. $N=56536$. Results were obtained using Monte Carlo
with 1000 trials.}

\label{fig:oavar_f_tot} 
\end{figure}

Figure (\ref{fig:oavar_f_tot}) presents a comparaison between the
Overlapping Allan variance computed in the frequency domain (F-OAVAR)
and the Total variance for three noise types : WHFM, FLFM and RWFM.
The upper plot depicts the \emph{edf} ratio computed using Monte Carlo
simulations with 1000 trials. we notice that the \emph{edf} of the
F-OAVAR and the Total variance are nearly identical. The lower plot
depicts the bias defined by $Bias=100\times(1-\sqrt{\textrm{F-OAVAR}/\textrm{Totvar}})$.
The bias of the F-OAVAR with respect to the Total variance is less
than 10\%. 

\begin{figure}
\centering \includegraphics[width=1\linewidth]{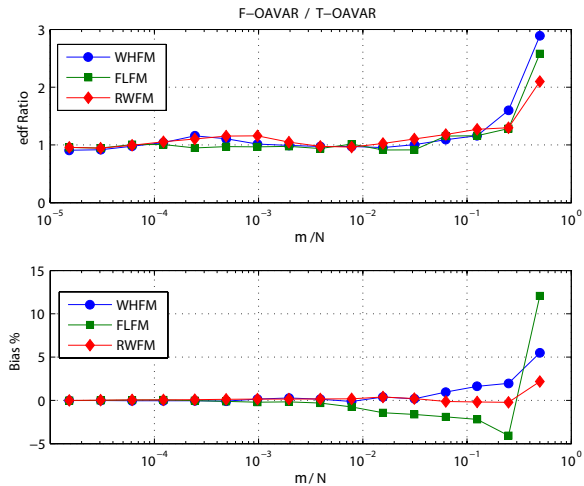} 

\caption{Comparaison of the F-OAVAR and the T-OAVAR for three noise types.
The upper plot depicts the edf ratio and the lower plot depicts the
bias. $N=56536$. Results were obtained using Monte Carlo with 1000
trials.}

\label{fig:oavar_f_t} 
\end{figure}

In the same way, figure (\ref{fig:oavar_f_t}) presents a comparaison
between the Overlapping Allan variance computed in the frequency domain
and the classical Overlapping Allan variance computed in the time
domain. The upper plot shows that the F-OAVAR \emph{edf }is two to
three times higher than the \emph{edf} of the T-OAVAR for the higher
$\tau$ value ($\tau=N/2$) .The lower plot depicts the bias defined
by $Bias=100\times(1-\sqrt{\mathrm{\textrm{F-OAVAR}}/\mathrm{\textrm{T-OAVAR}}})$.

\begin{figure}
\centering \includegraphics[width=1\linewidth]{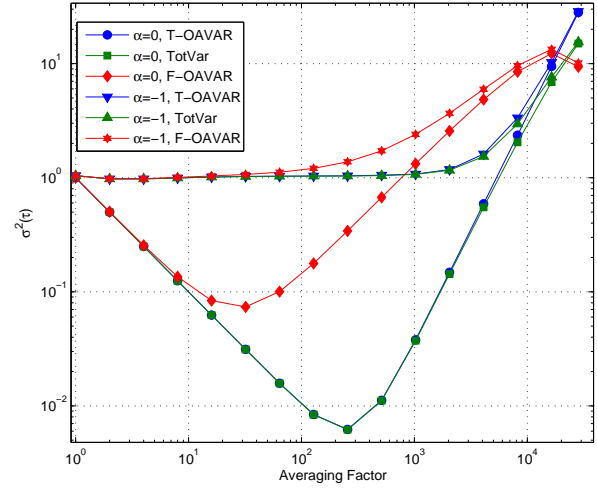} 

\caption{The T-OAVAR, the Total variance and the F-OAVAR for a White frequence
noise ($\alpha=0$) and a Ficker frequency noise ($\alpha=-1$) .
A linear frequency drift was added to the noise sequences of length
$N=56536$ (Monte Carlo trials = 1000). }

\label{fig:oavar_drift} 
\end{figure}

Figure (\ref{fig:oavar_drift}) shows the Total Variance, the Overalpping
Allan Variance computed in the time domain (T-OAVAR) and in the frequency
domain (F-OAVAR) for a White frequence noise and a flicker noise with
a linear frequency drift. The added linear drifts is equal to $D(t)=15t$
. Like the the Total variance and the classical Allan variance, the
F-OAVAR does not cancel the linear drift. We can notice also that
the F-OAVAR for a linear drift varies as $\tau$ , while the Total
variance and the T-OAVAR vary as $\tau^{2}$.

\begin{figure}
\centering \includegraphics[width=1\linewidth]{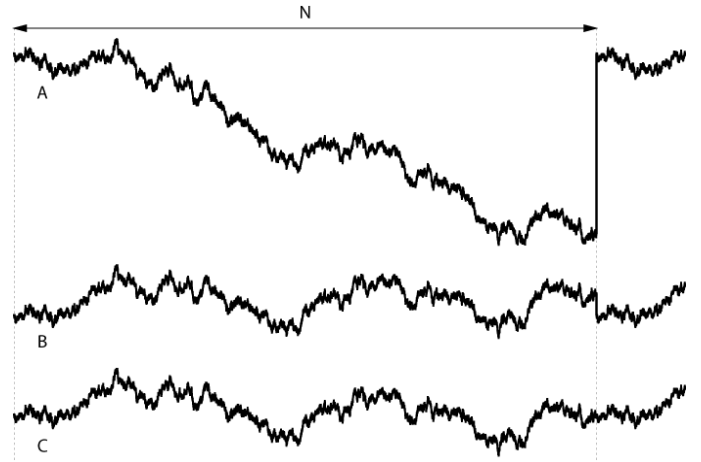}

\caption{Random Walk Frequency Noise sequence: rough (A), drift removed (B)
and circularized (C).}

\label{fig:circul} 
\end{figure}

Unfortunately, the last result shows that the computation of OAVAR
in the frequency domain presents a severe drawback: it is unable to
discriminate between a linear frequency drift and a $f^{-2}$ frequency
noise (random walk FM). This effect is due to the assumption of periodicity
of the sequence implicitely induced by the use of the FFT algorithm.
Figure \ref{fig:circul}-A shows that connecting the last sample to
the first one may induce a high edge, altering the variance measurements.
So we decided to process the frequency deviation sequence with 2 different
ways: 
\begin{itemize}
\item by removing the linear drift of this sequence (see figure \ref{fig:circul}-B;
let us notice that there is still an edge at the end of the sequence).
The removed line is estimated by a least squares fit of the data sequence
to a line. 
\item by circularizing the sequence (see figure \ref{fig:circul}-C), i.e.
by removing the linear drift in such a way that the last sample of
the residuals is equal to the first one. Denoting by $D(t)=a\cdot t+b$
the drift we have to substract from the sequence, the linear coefficient
$a$ is then: \begin{equation}
a=\frac{y_{N}-y_{1}}{t_{N}-t_{1}}\end{equation}
 and the constant term $b$ may be choosen equal to 0 since OAVAR
is not sensitive to additive constants. 
\end{itemize}

It is worth recalling that Figure (\ref{fig:circul}) shows the side
effect of periodization (induced by multiplication in the discret
frequency domain) of a sequence without processing, after a line removal,
and after circularization. But when computing the frequency domain
variances we don't realize any extension of data manually as done
in the computation of the Total variance.

Table (\ref{tab:deglib_fm2}) compares the \emph{edf }of the OAVAR
for a Random Walk Frequency Noise computed after these processings.
The best estimates are obtained by using the circularized sequence
since the \emph{edf }of the estimates are higher than for for the
sequence after removing a linear frequency drift. Thus, the edf of
the last estimate ($\tau=N/2$) is 2 times higher than the one of
the estimate obtained in the time domain. This means that this estimate
provides a noise level estimation $\sqrt{2}$ times sooner than the
estimate computed in the time domain (e.g. 265 days instead of 1 year)
with the same accuracy.

\begin{table}
\centering 

\caption{Comparison of the equivalent degrees of freedom of the time OAVAR
estimates and the spectral OAVAR estimates rough, after removing a
linear drift and after circularizing the sequence for a Random Walk
Frequency Noise sequence of length $N=65536$.}

\label{tab:deglib_fm2} \begin{tabular}{|r|r|r|r|r|}
\hline 
$\mathbf{\tau}$  & Time OAVAR  & \multicolumn{3}{|c|}{Spectral OAVAR}\tabularnewline
 &  & rough  & without drift  & circularized \tabularnewline
\hline 
1  & 68540  & 65660 & 39 & 56735\tabularnewline
2  & 35289 & 33269 & 39 & 27589\tabularnewline
4  & 15498  & 15410 & 39 & 13009\tabularnewline
8  & 7324  & 7725 & 38 & 6392\tabularnewline
16  & 3621  & 3997 & 38 & 3258\tabularnewline
32  & 1812  & 2091 & 37 & 1737\tabularnewline
64  & 900 & 1040 & 36 & 860\tabularnewline
128  & 455  & 477 & 35 & 436\tabularnewline
256  & 225  & 219 & 34 & 224\tabularnewline
512  & 110 & 106 & 31 & 109\tabularnewline
1024  & 52 & 54 & 25 & 50\tabularnewline
2048  & 25 & 28 & 17 & 23\tabularnewline
4096  & 12 & 14 & 10 & 11\tabularnewline
8192  & 5.3  & 6.7 & 4.8 & 5.3\tabularnewline
16384  & 2.4  & 3.1 & 2.0 & 2.6\tabularnewline
32768  & 1.0  & 2.0 & 1.5 & 2.1\tabularnewline
\hline
\end{tabular}
\end{table}

\begin{figure}
\centering \includegraphics[width=1\linewidth]{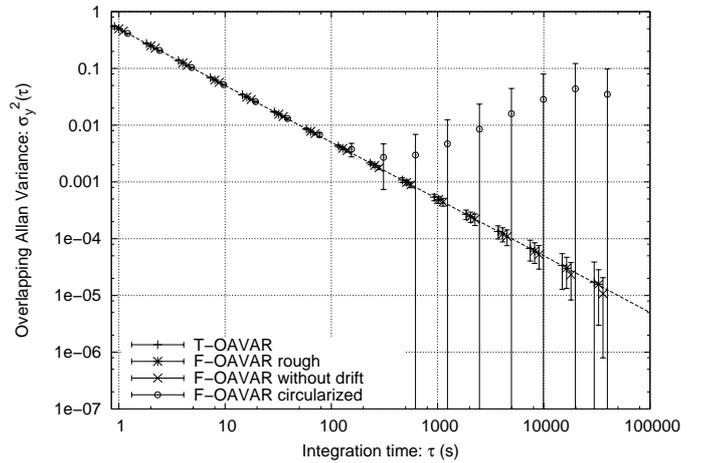} 

\caption{OAVAR for a White Frequency Noise sequence of length $N=65536$ computed
in the time domain and in the frequency domain, rough, after removing
the linear frequency drift and after circularizing the sequence.}

\label{fig:avar_blc} 
\end{figure}

However, applying the circularization processing to another type of
noise induced is a bias that has the same characteristic as a linear
frequency drift on an Allan variance plot. Beside the $\tau^{-1}$
behaviour characteristic of a white FM, figure \ref{fig:avar_blc}
exhibits the $\tau$ signature of a linear frequency drift in the
Allan variance curve of the circularized sequence. Let us also notice
the very long errorbars of the circularized sequence estimates. Therefore,
the circularization process cannot be used in a real frequency deviation
sequence which contains always different types of noise. Thus, we
recommand to apply the spectral OAVAR over the residuals of a frequency
deviation sequence, after removing the linear frequency drift. For
a random walk FM, the estimate of OAVAR computed in the frequency
domain after drift removal has an \emph{edf }1.5 times higher than
the classical time domain OAVAR. It means that spectral OAVAR after
drift removal is able to measure the random walk level of a sequence
$\sqrt{1.5}$ times sooner than time OAVAR (e.g. 300 days instead
of 1 year).

\begin{figure}
\centering \includegraphics[width=1\linewidth]{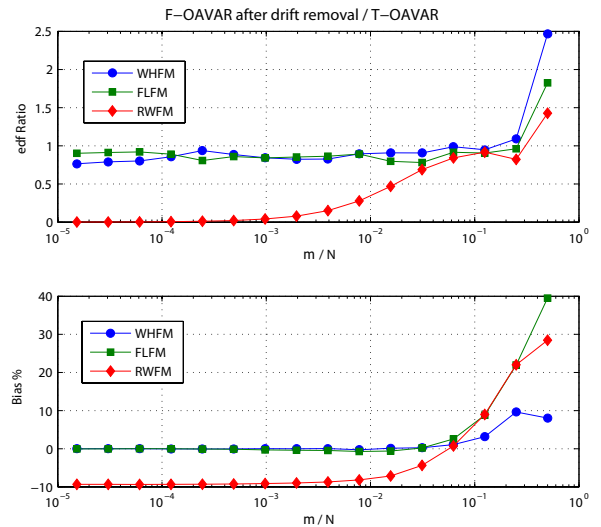} 

\caption{Comparaison of the F-OAVAR computed after drift removal from noise
sequences by least squares fit and the classical T-OAVAR for three
noise types. The upper plot depicts the \emph{edf} ratio and the lower
plot depicts the bias. $N=56536$. Results were obtained using Monte
Carlo with 1000 trials.}

\label{fig:oavar_f_dr} 
\end{figure}

Figure (\ref{fig:oavar_f_dr}) compares the F-OAVAR variance computed
after linear drift removal by least squares fit and the classical
T-OAVAR variance. As shown in Table (\ref{tab:deglib_fm2}) the upper
plot shows that the edf of the F-OAVAR after drift removal for a Random
Walk noise is less than the edf of the T-OAVAR for small $\tau$ values.
The lower plot shows that the F-OAVAR presents a bias of -10\% for
Random Walk noise. This bias can be explained by the fact the drift
removal from a Random Walk sequance alters the spectrum of the noise
at all the frequency values because a Random Walk contains a kind
of linear drift feature intrinsicly. 

Let us remember that for a sequence without random walk FM (for atomic
clocks), OAVAR computed in the frequency domain may be used directly
and is more accurate than OAVAR computed in the time domain.

\subsection{OHVAR}

\begin{figure}
\centering \includegraphics[width=1\linewidth]{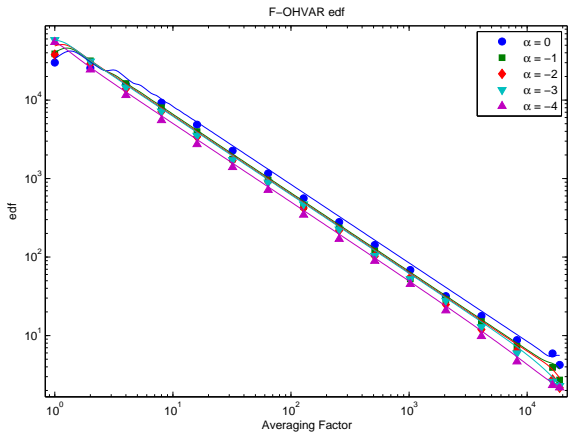} 

\caption{F-OHVAR \emph{edf} for five noise types ($\alpha$ from -4 to 0) for
sequences of length $N=65536$. The continous lines represent the
theoretical \emph{edf} computed by equation (\ref{eq:B17}). The symbols
represent the \emph{edf }obtained by Monte Carlo simulation with 1000
trials.}

\label{fig:hvar_edf1} 
\end{figure}

Figure (\ref{fig:hvar_edf1}) depicts the \emph{edf} of the Overlapping
Hadamard variance computed in the frequency domain F-OHVAR. It shows
a very good agreement between the theoretical \emph{edf }formula of
equation (\ref{eq:B17}) and the \emph{edf }obtained by Monte Carlo
simulations.

\begin{figure}
\centering \includegraphics[width=1\linewidth]{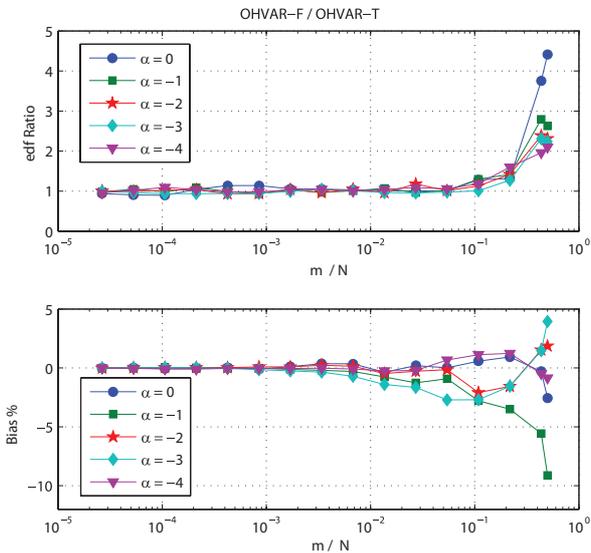} 

\caption{Comparaison of the F-OHVAR and the T-OHVAR for five noise types. The
upper plot depicts the \emph{edf} ratio and the lower plot depicts
the bias. $N=56536$. Results were obtained using Monte Carlo with
1000 trials.}

\label{fig:hvar_edf2} 
\end{figure}

Figure (\ref{fig:hvar_edf2}) shows that edf of the OHVAR estimator
in the frequency domain is 2 to 4.5 higher than the edf of the classical
OHVAR for the higher $\tau=N/3$ value. The lower plot depicts the
bias defined by $Bias=100\times(1-\sqrt{\mathrm{\textrm{F-OHVAR}}/\mathrm{\textrm{T-OHVAR}}})$.
It is less than 10\% for the five noise types and for all the $\tau$
values.

The Hadamard variance is not sensitive to linear frequency drifts.
However, computing OHVAR in the frequency domain by using a FFT assumes
also the periodicity of the sequence and may induce a high edge by
connecting the last sample to the first one (see figure \ref{fig:circul}-A).
We performed then the same processings as previously in order to compare
the effects of the drift removal and of the circularization of the
sequence. For OHVAR also, the circularization should not be recommanded
for processing frequency deviation sequences because it is only useful
for noises with $\alpha\leq-2$ and it degrades the variance estimates
for the noises with $\alpha>-2$ . On the other hand, the drift removal
by substracting the best least squares line from the data gives good
results for noises with $\alpha>-2$ . Hence, it is better to use
the F-OHVAR directly without preprocessing in order to get better
statistics than the T-OHVAR if the data does not contain a linear
drift.

\subsection{MAVAR}

\begin{figure}
\centering \includegraphics[width=1\linewidth]{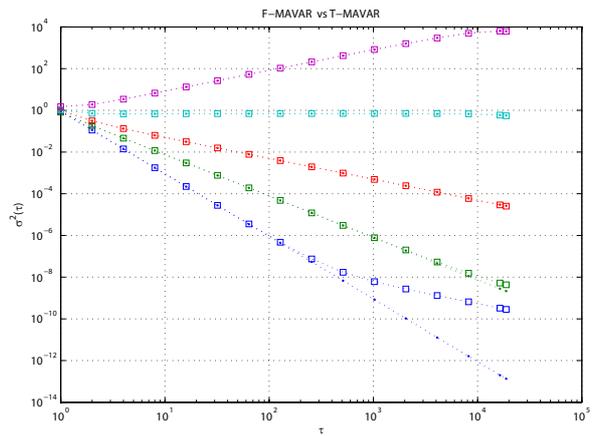} 

\caption{F-MAVAR and T-MAVAR for five noise types ($\alpha$ from -2 to +2)
for sequences of length $N=65536$. The squares represent the F-MAVAR
values and the dots represent the T-MAVAR values. Monte Carlo simulation
with 1000 trials.}

\label{fig:mavar_t_f} 
\end{figure}

Figure (\ref{fig:mavar_t_f}) shows a comparaison of the modified
Allan variance computed in the frequency domain (F-MAVAR) and in the
time domain (T-MAVAR) for five noise types with $\alpha$ from -2
to +2. We can notice clearly a huge bias of the F-MAVAR for $\alpha=+2$
.

For this reason, the use of MAVAR computed in the frequency domain
should be avoided.

\section{Conclusion}

We have presented a filter approach to analyze the different known
frequency stability variances. Using this approach we derived formulae
in the time domain identical to those known in the literature. We
also demonstrated for the first time that the computation of these
variances can be done in the frequency domain using a Discrete Fourier
Transform of the studied signals. Such a computation provides estimates
with better accuracy than the ones computed in the time domain, allowing
the measurement of the low frequency noise levels sooner, i.e. with
a shorter sequence. This advantage is particularly useful for studying
the long term stability of atomic clocks. However, in the presence
of linear drift, the periodicity of the sequence implicitely assumed
by the use of the FFT algorithm may induce edges which degrade variance
measurements if a random walk FM is present in the sequence. We have
demonstrated that, in this case, we must first remove the linear frequency
drift on a sequence before to compute a variance in the frequency
domain. Our work has proved that OAVAR computed in the frequency domain
is the estimator which gives the quickest low frequency noise level
(9 month instead of 1 year). New estimators improving these characteristics
with a more simple transfer function will be described in another
paper \cite{part_II}.

\section*{Appendix : Equivalence of the Discrete-Time and the Continuous-Time
variances }

\label{sect:AppxA} We have assumed $T=1$ in (\ref{eq:42}). Without
this assumption the variance ${{\sigma}_{{u}}^{2}\left(m\right)}$could
be written using (\ref{eq:sigy_sy}): \begin{eqnarray}
{\sigma}_{{u}}^{2}\left(m\right) & = & T\int_{-\frac{1}{2T}}^{+\frac{1}{2T}}S_{{u}}^{TS}(f)df\label{eq:43}\\
 & = & T\frac{2^{2n}}{c_{n}^{2}m^{2}}\int_{-\frac{1}{2T}}^{+\frac{1}{2T}}\frac{\sin^{2n+2}\left(\pi fmT\right)}{\sin^{2}\left(\pi{fT}\right)}S_{y}^{TS}\left(f\right)df.\nonumber \end{eqnarray}

Using expression (\ref{eq:PSDper}) of ${S_{y}(f)}$ in (\ref{eq:43})
we can write: \begin{eqnarray}
\sigma_{u}^{2}(m) & = & \frac{2^{2n}}{c_{n}^{2}m^{2}}\int_{-\frac{1}{2T}}^{+\frac{1}{2T}}\left\{ \frac{\sin^{{2n+2}}\left(\pi{fmT}\right)}{\sin^{2}\left(\pi{fT}\right)}\right.\label{eq:a1}\\
 &  & \left.\times\sum_{n}S_{Y}^{TS}\left(f-nf_{s}\right)\frac{\sin^{2}\left[\pi T\left(f-nf_{s}\right)\right]}{\left[\pi T\left(f-nf_{s}\right)\right]^{2}}\right\} df.\nonumber \end{eqnarray}

{The sine functions outside the sum sign are periodic, they can be
passed inside the sum sign. Doing this and making the variable change
} ${\nu=f-nf_{s}}${ we can write:} \begin{eqnarray}
\sigma_{u}^{2}(m) & = & \frac{2^{{2n}}}{c_{{n}}^{2}m^{2}}\sum_{k=-\infty}^{+\infty}{\int_{-\frac{1}{2T}-\frac{n}{T}}^{+\frac{1}{2T}-\frac{n}{T}}}\left[\frac{\sin^{{2n+2}}\left(\pi\nu mT\right)}{\sin^{2}\left(\pi\nu T\right)}\right.\nonumber \\
 &  & \left.\times\frac{\sin^{2}\left(\pi T\nu\right)}{\left(\pi T\nu\right)^{2}}S_{Y}^{TS}(\nu)\right]d\nu\label{eq:45}\end{eqnarray}
 where we have interchanged the sum sign and the integration symbol.

{ Equation (\ref{eq:45}) simplifies to:} \begin{eqnarray}
{\sigma}_{{u}}^{2}\left(m\right) & = & \frac{2^{{2n}}}{c_{{n}}^{2}}\int_{-\infty}^{+\infty}\frac{\sin^{{2n+2}}\left(\pi\nu mT\right)}{\left(\pi mT\nu\right)^{2}}S_{Y}^{TS}(\nu)d\nu\nonumber \\
 & = & {\sigma}_{{U}}^{2}\left(\tau\right)_{{(n)}}|_{{\tau={mT}}}.\label{eq:46}\end{eqnarray}

{The only difference between (\ref{eq:46}) and (\ref{eq:varUspec})
is the integration bounds. In equation (\ref{eq:varUspec}), } ${S_{Y}\left(f\right)}${
is the single-sided PSD while } ${S_{Y}^{TS}\left(u\right)}${ in
(\ref{eq:46}) is the two-sided PSD. We conclude that (\ref{eq:42})
and (\ref{eq:varUspec}) represent the same variance.

\bibliographystyle{IEEEtran}
\bibliography{papier_Alaa}

\begin{thebibliography}{10}
\providecommand{\url}[1]{#1}
\csname url@rmstyle\endcsname
\providecommand{\newblock}{\relax}
\providecommand{\bibinfo}[2]{#2}
\providecommand\BIBentrySTDinterwordspacing{\spaceskip=0pt\relax}
\providecommand\BIBentryALTinterwordstretchfactor{4}
\providecommand\BIBentryALTinterwordspacing{\spaceskip=\fontdimen2\font plus
\BIBentryALTinterwordstretchfactor\fontdimen3\font minus
  \fontdimen4\font\relax}
\providecommand\BIBforeignlanguage[2]{{%
\expandafter\ifx\csname l@#1\endcsname\relax
\typeout{** WARNING: IEEEtran.bst: No hyphenation pattern has been}%
\typeout{** loaded for the language `#1'. Using the pattern for}%
\typeout{** the default language instead.}%
\else
\language=\csname l@#1\endcsname
\fi
#2}}

\bibitem{totvar}
C.~A. Greenhall, D.~A. Howe, and D.~B. Percival, ``Total variance, an estimator
  of long-term frequency stability,'' \emph{IEEE Transactions on Ultrasonics,
  Ferroelectrics and Frequency Control}, vol. UFFC-46, no.~5, pp. 1183--1191,
  September 1999.

\bibitem{allan}
D.~W. Allan, ``Statistics of atomic frequency standards,'' \emph{Proceedings of
  the IEEE}, vol.~54, pp. 221--230, February 1966.

\bibitem{barnes}
J.~A. Barnes, A.~R. Chi, L.~S. Cutler, D.~J. Healey, D.~B. Lesson, T.~E.
  McCunigal, J.~A. Mullen, W.~L. Smith, R.~L. Sydnor, R.~Vessot, and G.~M.~R.
  Winkler, ``Characterization of frequency stability,'' \emph{IEEE Transactions
  on Instrumentation and Measurement}, vol. IM-20, pp. 105--120, May 1971.

\bibitem{linchi}
W.~C. Lindsey and C.~M. Chie, ``Theory of oscillator instability based upon
  structure function,'' \emph{Proceedings of the IEEE}, vol.~64, pp.
  1652--1666, December 1976.

\bibitem{modallan}
D.~Allan and J.~A. Barnes, ``A modified ``allan variance'' with increased
  oscillator characterization ability,'' in \emph{Proceedings of the
  35$^\textrm{\footnotesize st}$ Annual Frequency Control Symposium}, Fort
  Monmouth (NJ, USA), May 1981, pp. 470--475.

\bibitem{rutman}
J.~Rutman, ``Characterization of phase and frequency instabilities in precision
  frequency sources: fifteen years of progress,'' \emph{Proceedings of the
  IEEE}, vol.~66, no.~9, pp. 1048--1075, September 1978.

\bibitem{roddier}
F.~Roddier, \emph{Distributions et transformation de Fourier}.\hskip 1em plus
  0.5em minus 0.4em\relax Paris: McGraw-Hill, 1978.

\bibitem{papoulis}
A.~Papoulis, \emph{Probability, Random Variables, and Stochastic Processes},
  \mbox{3$^\textrm{\tiny rd}$}~ed.\hskip 1em plus 0.5em minus 0.4em\relax New
  York: McGraw Hill, 1991.

\bibitem{howe}
D.~A. Howe, R.~L. Beard, C.~A. Greenhall, F.~Vernotte, W.~J. Riley, and T.~K.
  Peppler, ``Enhancements to \mbox{GPS} operations and clock evaluations using
  a "total" hadamard deviation,'' \emph{IEEE Transactions on Ultrasonics,
  Ferroelectrics, and Frequency Control}, vol. UFFC-52, no.~8, pp. 1253--1261,
  August 2005.

\bibitem{vernotte02c}
F.~Vernotte, ``Application of the moment condition to noise simulation and to
  stability analysis,'' \emph{IEEE Transactions on Ultrasonics, Ferroelectrics,
  and Frequency Control}, vol. UFFC-49, no.~4, pp. 508--513, April 2002.

\bibitem{hadamard}
R.~Baugh, ``Frequency modulation analysis with the hadamard variance,'' in
  \emph{Proceedings of the 25$^\textrm{\footnotesize st}$ Annual Frequency
  Control Symposium}, June 1971, pp. 222--225.

\bibitem{picinbono}
E.~Boileau and B.~Picinbono, ``Statistical study of phase fluctuations and
  oscillator stability,'' \emph{IEEE Transactions on Instrumentation and
  Measurement}, vol. IM-25, no.~1, pp. 66--75, March 1976.

\bibitem{Walter}
T.~Walter, ``A multi-variance analysis in the time domain,'' \emph{24th Annual
  Precise Time and Time Interval (PTTI) Meeting}, pp. 413--424, 1992.

\bibitem{percival}
D.~B. Percival and A.~T. Walden, \emph{Wavelet Methods for Time Series
  Analysis}, ser. Cambridge Series in Statistical and Probabilistic
  Mathematics.\hskip 1em plus 0.5em minus 0.4em\relax Cambridge: Cambridge
  University Press, 2000.

\bibitem{metro98}
F.~Vernotte, G.~Zalamansky, and E.~Lantz, ``Time stability characterization and
  spectral aliasing. \mbox{P}art \mbox{I}: \mbox{A} time domain approach,''
  \emph{Metrologia}, vol.~35, no.~5, pp. 723--730, December 1998.

\bibitem{vernotte98b}
------, ``Time stability characterization and spectral aliasing. \mbox{P}art
  \mbox{II}: \mbox{A} frequency domain approach,'' \emph{Metrologia}, vol.~35,
  no.~5, pp. 731--738, December 1998.

\bibitem{percival2}
D.~B. Percival, ``Characterization of frequency stability: Frequency-domain
  estimation of stability measures,'' in \emph{Proceedings of the IEEE, VOL.
  79, NO. 6}, July 1991, pp. 961--972.

\bibitem{Chang}
P.~C. Chang, H.~M. Peng, and S.~Y. Lin, ``Allan variance estimated by phase
  noise measurements,'' \emph{36th Annual Precise Time and Time Interval (PTTI)
  Meeting}, pp. 165--172, 2004.

\bibitem{these_vernotte}
F.~Vernotte, ``Stabilit\'e temporelle des oscillateurs : nouvelles variances,
  leurs propri\'et\'es, leurs applications,'' PhD thesis, order N\# 199,
  Universit\'e de Franche-Comt\'e, Observatoire de Besan\c{c}on, February 1991.

\bibitem{howe2}
D.~A. Howe, D.~W. Allan, and J.~A. Barnes, ``Properties of signal sources and
  measurement methods,'' in \emph{Proceedings of the 35$^\textrm{\footnotesize
  st}$ Annual Frequency Control Symposium}, Fort Monmouth (NJ, USA), May 1981,
  pp. A1--A47.

\bibitem{allan_eftf}
D.~W. Allan, ``Time and frequency metrology: current status and future
  considerations,'' \emph{5th EFTF, 1-9, Besançon}, 1999.

\bibitem{fft}
J.~W. Cooley and J.~W. Tukey, ``An algorithm for the machine calculation of
  complex fourier series,'' \emph{Math. Comput.}, vol.~19, no.~90, pp.
  297--301, April 1965.

\bibitem{lesage}
P.~Lesage and C.~Audoin, ``Characterization of frequency stability: uncertainty
  due to the finite number of measurements,'' \emph{IEEE Transactions on
  Instrumentation and Measurement}, vol. IM-22, pp. 157--161, June 1973, see
  also corrections published in 1974, March and 1976, September.

\bibitem{part_II}
A.~Makdissi, F.~Vernotte, and E.~Declercq, ``Stability variances: New variances
  in the frequency domain,'' To be published.

\end{thebibliography}
 
\end{document}